\documentclass[a4paper,fleqn,usenatbib,useAMS]{mnras}
\usepackage{graphicx}	
\usepackage{epstopdf}

\usepackage{amsmath}	

\usepackage{amssymb}	

\usepackage{txfonts}

\usepackage{float}

\newcommand{\hctn}{HC$_3$N}
\newcommand{\hcfn}{HC$_5$N}
\newcommand{\hcsn}{HC$_7$N}
\newcommand{\cts}{C$_3$S}
\newcommand{\um}{$\mu$m}
\newcommand{\cfh}{C$_4$H}
\newcommand{\cfht}{C$_4$H$_2$}
\newcommand{\chtcch}{CH$_3$CCH}

\newcommand{\kms}{$\mathrm{km~s^{-1}}$}

\newcommand{\altaffilmark}[1]{$^{#1}$}
\newcommand{\tablenotemark}[1]{$^{#1}$}
\newcommand{\tablenotetext}[2]{$^{#1}$#2}
\newcommand{\nodata}{}
\newcommand{\colhead}[1][]{#1}

\title[Carbon-Chain Molecules in Molecular Outflows and Lupus I]{Carbon-Chain Molecules in Molecular Outflows and Lupus I Region
--New Producing Region and New Forming Mechanism}

\author[Yuefang Wu et al.]{Yuefang Wu\altaffilmark{1,2}\thanks{E-mail: ywu@pku.edu.cn},
    Xunchuan Liu\altaffilmark{1,2},
	Xi Chen\altaffilmark{3,4}, Lianghao Lin\altaffilmark{1,5,6}, Jinghua Yuan\altaffilmark{7}, \newauthor
     Chao Zhang\altaffilmark{1,2,8},
     Tie Liu,\altaffilmark{9,10} Zhiqiang Shen\altaffilmark{3}, Juan Li\altaffilmark{3}, Junzhi Wang\altaffilmark{3}, \newauthor
        Sheng-Li Qin\altaffilmark{8}, Kee-Tae Kim\altaffilmark{9},
		 Hongli Liu\altaffilmark{11}, Lei Zhu\altaffilmark{7}, Diego Madones\altaffilmark{12,13}, \newauthor
          Natalia Inostroza\altaffilmark{14},  C. Henkel\altaffilmark{15,16,17}, Tianwei Zhang\altaffilmark{1,2},
		  Di Li\altaffilmark{7,18,19},  Jarken Esimbek\altaffilmark{17,20} \newauthor
          and Qinghui Liu\altaffilmark{3}\\
        $^1$Department of Astronomy, School of Physics, Peking University, 100871 Beijing, China \\
        $^2$Kavili Institute for Astronomy and Astrophysics, Peking University, 100871 Beijing, China \\
        $^3$Shanghai Astronomical Observatory, Chinese Academy of Sciences, Shanghai 200030, China \\
        $^4$Center for Astrophysics, GuangZhou University, Guangzhou 510006, China \\
        $^5$School of Astronomy and Space Sciences, University of Science and Technology of China, 96 Jinzhai Road, Hefei, 230026, China \\
        $^6$Purple Mountain Observatory and Key Laboratory of Radio Astronomy, Chinese Academy of Sciences, 8 Yuanhua Road, Nanjing, 210034, China \\
        $^7$National Astronomical Observatories, Chinese Academy of Sciences,
        Beijing 100101, China \\
		$^8$Department of Astronomy, Yunnan University,
			Kunming, 650091, China \\
		$^9$Korea Astronomy and Space Science Institute, 776 Daedeokdae-ro,
			Yuseong-gu, Daejeon 34055, Korea \\
        $^{10}$East Asian Observatory, 660 North A$\arcmin$ ohoku Place, Hilo, HI 96720, USA\\
		$^{11}$Department of Physics, The Chinese University of Hong Kong,
			Shatin, NT, Hong Kong SAR \\
        $^{12}$Department of Astronomy, University of Chile, Casilla 36-D, Santiago, Chile\\
        $^{13}$ Centre for Astrochemical Studies, Max-Planck-Institute for Extraterrestrial Physics, Giessenbachstrasse 1, 85748, Garching, Germany \\
        $^{14}$ N\'{u}cleo de Astroqu\'{i}micay Astrof\'{i}sica, Instituto de Ciencias Qu\'{i}micas Aplicadas, Facultad de Ingenier\'{i}a, Universidad Aut\'{o}noma de Chile Av.\\
         Pedro de Valdivia 425, Providencia, Santiago de Chile.\\
        $^{15}$Max-Planck Institut f\"{u}r Radioastronomie, Auf Dem H\"{u}gel 69, 53121 Bonn, Germany\\
        $^{16}$Astronomy Department, Faculty of Science, King Abdulaziz University, PO Box 80203, Jeddah, 21589, Saudi Arabia\\
        $^{17}$Xinjiang Astronomical Observatory, Chinese Academy of Sciences, 830011, Urumqi, China\\
        $^{18}$Key Laboratory of Radio Astronomy, Chinese Academy of Science, Nanjing, China\\
        $^{19}$University of Chinese Academy of Sciences, Beijing 100049, China\\
        $^{20}$Key Laboratory of Radio Astronomy, Chinese Academy of Sciences, Urumqi 830011, China}

\begin{document}
\label{firstpage}
\pagerange{\pageref{firstpage}--\pageref{lastpage}}
\maketitle

\begin{abstract}
	Using the new equipment of the Shanghai Tian Ma Radio Telescope, we have searched for carbon-chain molecules (CCMs) {\bf towards} five outflow
    sources and six Lupus I starless dust cores, including one region known to be
      characterized by
 warm carbon-chain chemistry (WCCC), Lupus I-1 (IRAS 15398-3359), and one TMC-1 like cloud, Lupus I-6 (Lupus-1A).
Lines of \hctn~$J=2-1$, \hcfn~$J=6-5$, \hcsn~$J=14-13$, $15-14$, $16-15$ and \cts~$J=3-2$ were detected in all the targets except in the outflow source L1660 and the starless dust core Lupus I-3/4.
The column densities of nitrogen-bearing species range from 10$^{12}$ to 10$^{14}$ cm$^{-2}$ and those of C$_3$S are  about 10$^{12}$ cm$^{-2}$.
Two outflow sources, I20582+7724 and L1221, could be identified as new carbon-chain--producing regions.
Four of the Lupus I dust cores are newly identified as early quiescent
and dark carbon-chain--producing regions similar to Lup I-6, which together with the WCCC source, Lup I-1,
indicate that  { carbon-chain-producing regions are popular in Lupus I which { can be regard as} a Taurus like molecular cloud
complex} in our Galaxy. The column densities of \cts~are larger than those
of \hcsn~in the three outflow sources I20582, L1221 and L1251A.
Shocked carbon-chain chemistry (SCCC) is proposed to explain the abnormal high abundances of \cts~compared with those of nitrogen-bearing CCMs.
Gas-grain chemical models support the idea that shocks can fuel the { environment} of
those sources with enough $S^+$ thus driving the generation of S-bearing CCMs.

\end{abstract}


\section{Introduction} \label{sec_intro}

	Carbon-chain molecules are important components of the interstellar gas.
    They are major players in hydrocarbon chemistry 
	and sensitive indicators of evolutionary states of molecular regions due to their wide range { of} masses and large permanent dipole moments  \citep{2001ApJ...558..693D,2011JChPh.135x4310I,2012MNRAS.419..238B}.  
	They can also trace dynamic processes in molecular regions including material
	infall \citep{2013MNRAS.436.1513F}.
After being first detected in { the} massive star-forming region Sgr B2 \citep{1971ApJ...163L..35T, 1976ApJ...205L.173A},
CCMs were found in interstellar { clouds} with different evolutionary states.
A number of small carbon-chain molecules such as C$_2$H, C$_3$H$_2$ and C$_3$H were detected in diffuse clouds \citep{1979Natur.278..722M,2000A&A...358.1069L,2012ApJ...753L..28L}.
These molecules were also found in photodissociation regions \citep{2004A&A...417..135T,2013A&A...557A.101G,2005A&A...435..885P}.
Their abundances were explained by a carbon-chain producing mechanism in which the photo-erosion of UV-irradiated large carbonaceous compounds could feed the ISM with small carbon clusters or molecules \citep{2004A&A...417..135T}.
CCMs have also been also detected in dense molecular cores.

In the early phase of stellar evolution, a number of cold and dark cores were found as abundant producing regions of CCMs.
Among them, TMC-1 is one of the best studied regions where sulfur-, oxygen- and deuterium-bearing CCM species were found \citep{1980BAAS...12..485L,1984Natur.310..125M,1988ApL&C..26..167I, 1997ApJ...483L..61B,2004PASJ...56...69K}.
The most recent one is HC$_5$O detected by  \citet{2017ApJ...843L..28M}. It is a good test source of chemical models of cold dark cores \citep{1984MNRAS.207..405M,2014MNRAS.437..930L}.
In \citeyear{1992ApJ...392..551S}, \citeauthor{1992ApJ...392..551S} studied a sample consisting of 27 starless dark cores and 22 star-forming cores.
\citet{2010ApJ...718L..49S} identified Lupus 1A (Lup I-6) as the ``TMC-1 like'' cloud in the Lupus region.
 Such cores are in an early stage of chemical evolution 
 \citep{1992ApJ...392..551S,2006ApJ...646..258H}.

In star-forming cores,  CCMs become deficient \citep{2008ApJ...672..371S}.  Abundances of S-bearing species such as C$_2$S are  much lower than those in cold and dark cores \citep {1992ApJ...392..551S}.
In fact, depletions of S-bearing species start much earlier in the densest regions.
In the starless core L1544,
observations with BIMA { have} revealed that C$_2$S emission \citep{1999ApJ...518L..41O} shows a shell like structure,  
while at the center N$_2$H$^+$ emission is concentrated \citep{1999ApJ...513L..61W}. {
Recently \citet{2018MNRAS.478.5514V} detected 21 S-bearing species towards L1544 and found that a strong depletion was needed to explain the results}.
\citet{1992ApJ...392..551S} found that C$_\mathrm{n}$S and C$_\mathrm{n}$H (n=1-3) reached their { maximum} at A$_V$ $\leq$1.2 in the gas phase.
Among the sources they selected, protostellar cores  such as L1489, L1551 and L1641N were not detected or only marginally detected in C$_2$S and C$_3$S. { However among 16 deeply embedded low mass protostars,  C$_2$S and \cts~ were detected in 88$\%$ and 38$\%$ of the targets respectively \citep{2018ApJ...863...88L}}.

  High excitation lines of carbon-chain molecules were detected toward low-mass star-forming cores.
Lines such as  \cfh~(N=9-8), \chtcch~(J=5-4, K=2) and \cfht~$J=10_{0,10}-9_{0,9}$ were detected in L1527 and IRAS 15398-3359 \citep{2008ApJ...672..371S, 2009ApJ...697..769S}.
The CCM chemistry in these regions is different from that in early cold and dark cores,
thus a new mechanism called warm carbon-chain chemistry (WCCC) was suggested by \citet{2008ApJ...672..371S}.
High excitation linear hydrocarbons in these regions are formed by reactions of CH$_4$, { i.e. from} molecules that have been sublimated from warm dust grains \citep{2008ApJ...672..371S,2008ApJ...681.1385H}.
Another possible reason for the high abundance of CCMs in the WCCC sources is the shorter time scale of prestellar collapse in these cores compared with those in other protostellar cores, which results in the survival of CCMs \citep{2008ApJ...672..371S,2009ApJ...697..769S}.
Besides high-excitation species, N-bearing CCMs are also abundant in WCCC sources.
Lines of \hctn, \hcfn, \hcsn~and even HC$_9$N were detected in L1527.
\hcfn- J= 32-31 with very high excitation energy (67.5 K) was detected in { a} second WCCC source IRAS 15398-3359 \citep{2009ApJ...697..769S}.
However, the S-bearing species in WCCC sources are less abundant than in common star-forming cores \citep{2008ApJ...672..371S,1992ApJ...392..551S}.

Despite this notable progress, CCMs remain not yet fully understood.
How do the emissions of CCMs evolve as they undergo longer periods of heating by protostellar sources when compared to WCCC sources?
Are the emissions of CCMs affected by dynamical feedback from the protostars such as molecular outflows and jets?
Are the N-bearing species still abundant there? Will the S-bearing species be quenched?
For the cores at early stellar phase, so far the carbon-chain producing regions detected are mainly located in the Taurus  molecular complex \citep{2009ApJ...699..585H,1992ApJ...392..551S,1981ApJ...244...45S}.
Are there other CCM rich molecular complexes similar to the Taurus complex in our Galaxy?


    In this paper, we present observations of \hctn, \hcfn~and \hcsn~
	as well as \cts~ in the  15.2 to 18.2 GHz frequency range toward five molecular outflow sources and six Lupus I starless dust
	cores to investigate their emissions and production mechanisms. The Lup I cores were numbered with the numbers of the dust cores of \citet{2015A&A...584A..36G}.
A known WCCC core, IRAS 15398-3359 (Lup I-1), and the TMC-1 like core, Lupus 1A (Lup1-6), are included to explore their CCM emissions 
and to compare the derived properties with those obtained from other samples.
The observed sources
	are listed in \autoref{table_source}. Parameters related to the observed lines are given in \autoref{table_freq}, which are taken from the ``Splatalogue'' molecular database \footnote{\url{www.splatalogue.net}}.
Our observation is introduced in \autoref{sec_obs}.
In { Sects.} 3 and 4
we present results and
	discussions. \autoref{sec_summary} provides a summary.

\section{Observation} \label{sec_obs}
The observations were carried out with the Tian Ma Radio Telescope (TMRT) of the Shanghai Observatory.
	The TMRT is a newly built 65-m diameter fully steerable
radio telescope located in the western outskirts of Shanghai \citep{ 2016ApJ...824..136L}.
 The pointing accuracy is better than 10\arcsec,
and the main beam efficiency is 0.60 { in}
	the  12-18 GHz band
 \citep{2015AcASn..56...63W,2016ApJ...824..136L}.
The front end is
	a cryogenically cooled receiver covering the frequency range of
	11.5$-$18.5 GHz.
    An FPGA-based spectrometer
	based upon the design of the Versatile GBT Astronomical Spectrometer
	(VEGAS) was employed as the Digital backend system (DIBAS)
	\citep{2012AAS...21944610B}. For molecular line observations,
DIBAS supports a variety of observing modes, including 19
single-subband modes and 10 eight-subband modes.
	The center frequency of each subband is tunable
to an accuracy of 10 kHz.
For our observations mode 22 was adopted.
 Each of the eight subbands in two banks (Bank A and Bank B)
has a bandwidth 23.4 MHz and 16384 channels.
The velocity resolution is { by} a little larger than the channel spacing which  is 0.028 km s$^{-1}$ in the 15 GHz band and 0.023 km s$^{-1}$ in the 18 GHz band respectively.
{ The calibration uncertainty is 3 percent \citep{2015ChA&A..39..394W}.}

	As mentioned above, among the observed sources there are five outflow sources
	and six Lupus I starless dust cores. The numbers of the Lupus I cores labeled by \citet{2015A&A...584A..36G}
	have been adopted. Lupus is abbreviated as Lup. Source names, alternative
	names, positions, distances, references and observational notes are
	listed in columns 1 to 7 of \autoref{table_source}.
	
    The measured carbon-chain molecular transitions are listed in \autoref{table_freq}.
    Columns 2-6 show the transitions covered
	by the 16 subbands, and their frequencies,
	the upper energies
	as well as the Einstein transition coefficients for spontaneous emission.

The half power beam widths (HPBWs) of the TMRT beam at our observed frequencies range from 52 to 60 arcsec, and are listed in the Column 6 of  \autoref{table_freq}.
\autoref{beams} shows regions covered within a single beam towards each source.
For outflow sources, the beam can cover the blue and red lobes of outflows either entirely or at least their main parts, including the driving objects, H$_2$ outflows and optical jets (also see Sect. \ref{sec_JS_Phy}).
The Herschel 250 $\mu$m images of the outflow sources have sizes less than or similar
to our beam size, and the beam can cover them too (\autoref{table_SED}). Our beam can also cover the Herschel
250 $\mu$m images of the Lupus cores entirely (\autoref{beams} and \autoref{table_SED}).

 	The  
 rms noise (Part II of \autoref{table_obs_par}) of IRAS 20582+7724 and L1221 (8 mK $-$ 21 mK)
     is much lower than the corresponding values of the remaining sources (30 mK $-$ 100 mK) because supplementary observations were made towards these two sources on Nov. 9, 2017 (\autoref{table_freq}).
 	For  observations in the 16-18 GHz range towards  cores Lup I-1 and Lup I-7/8/9, the rms noise is higher (100 mK $-$ 200 mK)
 	because they were measured on 2016 March 21 and March 24
 	under relatively poor weather conditions.

 	The package GILDAS including CLASS and GREG
 	\citep{2000ASPC..217..299G} was used to reduce the data and
    draw the spectra.

\section{Results} \label{sec_results}
	\subsection{ Detected lines}
	Lines in the observed band were detected
	in all our targets except towards the outflow source L1660 and the dust core Lup I-3/4.
	The spectral lines of \hctn~$J=2-1$ for the  detected  sources are shown in \autoref{fig_hctn}.
    Five hyperfine components of \hctn~$J=2-1$ were well resolved for all the detected sources.
	Panels (a)-(i) of \autoref{fig_spectra} present the remaining spectra of the observed
	transitions for each detected source.

    In IRAS 20582+7702 (hereafter I20582) and L1221, lines of \hctn~and \hcfn~were detected. It is the first time to detect CCMs in
	these two sources.
    Lines of \hcsn~were weak or not detected.
	Emission of \cts~$J=3-2$ was detected and turned out to be stronger than the emissions of the three rotation transitions of \hcsn~for these two sources.

	In L1251A, CCMs were { also reported} at higher frequency transitions by \citet{2011ApJ...730L..18C} but were detected { in} the observed band for the first time.
	From \autoref{fig_spectra}(c) one can see that the $J=6-5$ line
    of \hcfn~was detected and the hyperfine component $F=5-4$ was resolved. 
    The $J=14-13,~15-14$ and $~16-15$
	transitions of \hcsn~were detected.
    Emission of \cts~$J=3-2$ is stronger than emissions of the three transitions of \hcsn~too.

	The observed lines of Lup I-1 (IRAS 15398-3359) are the strongest among all the observed sources including outflows and dust cores of Lupus I.
    The only exception is
	the emission of \cts, which is slightly weaker than that of L1251A.
    Emissions of \hcsn~$J=14-13,~15-14,~16-15$ from this core  are all stronger than that of \cts~$J=3-2$.

	In the Lupus I region, besides outflow source Lup I-1, six starless cores were mapped
	in the dust continuum at 850 \um~\citep{2015A&A...584A..36G}.
	Lup I-6 was found as a ``TMC-1 like cloud'' and named as Lupus-1A by \citet{2010ApJ...718L..49S}. All the observed transitions are detected in the six cores
	except Lup I-3/4, which is therefore not displayed in { Figures 1 and 2}.
    Lup I-7/8/9 and Lup I-11 show strongest emission of \hctn~J=2-1 while Lup I-2 shows weakest.
	The \hcfn~hyperfine components
	$F=7-6,~6-5$, and $5-4$ are well resolved in Lup I-6, Lup I-7/8/9 and
	Lup I-11.
    The emission of \cts~is strongest in Lup I-11 and weakest in Lup I-6.
    However, emission of \cts~$J=3-2$  is weaker than that of the three rotation transitions of \hcsn~for all the five detected Lupus I cores.

Emissions of our searched lines were not detected in
		two sources, L1660 and Lup I-3/4.
       L1660 is an outflow source and possesses an H$_2$ jet \citep{1997A&A...324..263D}.
       There are no Herschel data within 70 arcmin of L1660. The reason for the non-detection of CCMs needs to be further examined.
Cores Lup I-3/4 are associated with a bright emission nebula (B77) and a reflection nebula (GN 15.42.0) within about 35\arcsec~\citep{1977A&AS...29...65B,2003A&A...399..141M}.
Its rather advanced evolutionary state and hot environment seem to be adverse to the production of CCMs.

\subsection{Line parameters}
All the resolved hyperfine structure (HFS) of detected spectral lines
was fitted with independent Gaussian { functions}. The observed parameters
including { Local Standard of Rest center velocity V$_\mathrm{lsr}$, peak temperature $T_\mathrm{MB}$,
 and full width to half maximum (FWHM) line width} as well as the integrated area are given in  Part I-IV of
\autoref{table_obs_par} respectively.

From \autoref{table_obs_par} as well as { Figures 1 and 2}, 
one can see that
the V$_\mathrm{LSR}$ of transitions of different molecules
	agrees with each other quite well.
The line widths of { the }outflow sources I20582 and L1221 are the widest, while the line
widths of { the} cores Lup I-7/8/9 and Lup I-11 are narrower, which are still with high ratios to the velocity resolution.
The main beam brightness temperatures T$_\mathrm{MB}$ of the \hctn~$J=2-1,~F=3-2$ lines are the highest among all the detected transitions for each source.
{ The emissions of this transition are quite strong in all the Lupus I starless cores. However, the outflow Lup I-1 has the strongest emission, while the emissions of the other three outflows  are all much weaker than those of Lup I-1 and the starless cores.}

The T$_\mathrm{MB}$ of this transition of the five Lupus I starless  dust cores ranges from  1.71 K (Lup I-2) to 4.10 K (Lup I-7/8/9).
The outflow Lup I-1, a WCCC source, has the highest T$_\mathrm{MB}$ of this transition (5.11 K) among the detected sources.
However, for the three outflow sources I20582, L1221 and L1251A, the T$_\mathrm{MB}$ values of \hctn~$J=2-1,~F=3-2$  are  much lower than those of Lup I-1 and the five Lupus I starless  dust cores.

 For the outflow source Lup I-1 and the 5 Lupus I starless  dust cores, three hyperfine components of \hcfn~$J=6-5$ were well or partially resolved.
While for the three outflow sources I20582, L1221 and L1251A,
the emissions of \hcfn~are weaker than those of the Lup I-1 and the 5 Lupus I starless  dust cores.
The T$_\mathrm{MB}$ of the three rotational transitions of \hcsn~ranges from 0.18 to 0.49 K for the Lupus cores.
For the three outflow sources the emissions of these lines are not or only marginally detected.

\cts~emission was detected in all our sources except L1660 and Lup I-3/4.
The  emissions of \cts~are weaker than those of all the N-bearing molecules in Lup I-1 and the five Lupus I starless  dust cores.
However, for the three outflow sources, the emission peaks of \cts~are higher than those of  \hcsn~$14-13$, $15-14$ and $16-15$.


\subsection{Column densities}
Column densities of the observed molecular species were calculated assuming local thermodynamic equilibrium (LTE) with the solution of the radiation transfer equation
  \citep {1991ApJ...374..540G,2015PASP..127..266M}.
	
	 	\begin{eqnarray}
	 		N= \frac{3k}{8\pi^3v}\frac{Q}{S_{ij}\mu^2}
              \frac{J(T_{ex})exp(\frac{E_{up}}{kT_{ex}})}{J(T_{ex})-J(T_{bg})} \frac{\tau}{1-e^{-\tau}}
	 	         \int T_r d\upsilon
        \end{eqnarray}
	 	\begin{eqnarray}
          J(T)=(exp^{\frac{hv}{kT}}-1)^{-1}
        \end{eqnarray}
	where $B$, $\mu$, and $Q$ are the rotational constant,
	the permanent dipole moment and the partition function adopted from  ``Splatalogue''
\footnotemark[1].

{ Excitation temperatures can be derived from the HFS fittings of \hctn~ $J=2-1$ (denoted as T$_{ex}$(HC$_3$N)) 
 with the fitting program in GILDAS/CLASS\footnote{\url{http://www.iram.fr/IRAMFR/GILDAS/doc/html/class-html}}.
Five hyperfine lines of HC3N J=2-1 are detected towards all target sources. 
Optical depths ($\tau$(HC$_3$N)) and excitation temperature (T$_{ex}$(HC$_3$N)) are listed in column 2 and 3 of \autoref{table_derived_hfs} respectively.
However, this method can not be applied to sources I20582 and L1251A, where the HFS of \hctn~indicates optically thin emission and the uncertainty introduced by the beam filling factor can not be ignored.}

 The observed HFS lines  of \hcfn~J=6-5, F=5-4, 6-5 and 7-6 are optically thin. \hcfn~ J=6-5, F=5-4, and 7-6 were not detected toward I20582 and L1221 while J=6-5, F=6-5 was not detected in L1251A, Lup I-2, and Lup I-5. The three rotation lines of \hcsn~J=14-13, 15-14, and 16-15 have similar E$_{up}$ { (Table 2)}. Furthermore the \hcsn~lines of J=14-13 in L1221, and J=15-14 in I20582 as well as
J=16-15 in L1221 do not have enough S/N. Therefore one can not derived excitation temperature from the observed molecular lines of \hcfn~ and \hcsn~.



Dust temperature ($T_\mathrm{d}$) as well as column densities of hydrogen molecules
are derived	from SED fitting of Herschel data at 70,
160, 250, 350, and 500 $\mu$m (See Appendix \ref{sec_appendix}) and are listed in \autoref{table_SED}.
For comparisons, dust parameters of TMC-1 
are also included.
T$_\mathrm{d}$ of Lup I-1 is 13.9 K (see \autoref{table_SED}), which is close to the molecular gas temperature of this source and of L1527 derived by \citet{2009ApJ...697..769S,2008ApJ...672..371S}.
 The dust temperatures T$_\mathrm{d}$ are also listed in column 4 of \autoref{table_derived_hfs} for the purpose of comparison.
 In  Lup I-7/8/9, T$_\mathrm{d}$  is lower than T$_{ex}$(HC$_3$N) by 4.3 K, i.e. 10.2 K versus 14.5 K (\autoref{table_derived_hfs}) which is the largest difference
 between these two temperature values of all the sources. This may be due to the fact that our Lup I -7/8/9 spectra are a combination of Lup I-7, Lup I-8 and Lup I-9 which are all inside our TMRT beam (see Table 3 of \citet {2015A&A...584A..36G}, and Figure A1). According to \citet{2015A&A...584A..36G} the dust temperatures of these three cores derived from Herschel data are all 13 K individually, which is close to our T$_{ex}$(HC$_3$N). The TMC-1 like cloud Lup I-6 is characterized by T$_{ex}$(HC$_3$N) $\sim$7.0 K and T$_\mathrm{d}$ $\sim$10 K, which are rather close to each other
 compared with an excitation temperature of 7.3$\pm$1.0 K given by \citet{2010ApJ...718L..49S} on the basis of C$_6$H and the dust temperature of 13 K given by \citet{2015A&A...584A..36G}.
The difference of these two values results in an error of derived column density related to the
estimation of T$_{ex}$, but the error is
not larger than 15 percent.

Thus, we assume in the following that T$_\mathrm{ex}$ equals the dust temperature (T$_\mathrm{d}$) under local thermodynamic
equilibrium (LTE) conditions \citep{2012ApJ...756...60S}.
 { The  column densities derived from each detected line are given in
Part I of \autoref{table_derived_par_detailed}} for all the sources.
One can see that for every species in each source, the column densities derived from different line components tend to be close to each.

 { Part II of \autoref{table_derived_par_detailed} lists}
the unweighed average of the column densities derived from different hyperfine components or rotation lines, which are adopted as the column density of a detected species.

 From  \autoref{table_derived_par_detailed}, one can see that among different species,
the column density of \hctn~is the highest. 
The WCCC source Lup I-1 shows the highest value. 
Lup I-5 and Lup I-11 also have column densities close to Lup I-1.
The three outflow sources I20582, L1221 and L1521A have the lowest \hctn~column densities.
The \cts~column densities of the WCCC source Lup I-1 and the five Lupus I cores  are all lower than column densities of their N-bearing species.
However, \cts~column densities of the three outflow sources are higher than their \hcsn~column densities.

\subsection{Abundance}
With the column densities of  molecules presented in \autoref{table_derived_par_detailed}, { abundances relative to $N$(H$_2$) (see \autoref{table_SED})} of detected species were obtained,
 which are given in \autoref{table_derived_par}. {\autoref{fig_abundance}} presents the changes of abundances among N-bearing LCCMs
	HC$_\mathrm{2n+1}$N (n=1-3) and between N-bearing species and \cts.

For N-bearing species HC$_\mathrm{2n+1}$N (n=1-3), the abundance decreases while n increases.
WCCC source Lup I-1 and the five Lupus I starless cores show the highest abundances.
Abundances of N-bearing species are lower in the other three outflow sources.

 For the molecule \cts,  the abundance in L1251A is the highest among all the detected sources.
The abundance of \cts~in I20582 or L1221 is also quite high compared with that of the TMC-1 like cloud Lup I-6.
The three outflow sources all have abundances of \cts~larger than that of \hcsn,
especially in the case of L1251A in which the abundance of \cts~is even comparable to that of \hcfn.



{\autoref {table_calculated_values}} 
presents the abundance ratios of all the species for all the sources.
 The ratios of x(\hctn)/x(\hcfn)
 for the Lup I starless cores and WCCC source Lup I-1 are all rather close to each other, with an average value of $\sim$3.8. However the ratios of the other three outflow sources are 6.8 on the average and higher than those of the starless cores and the Lup I-1 outflow source.
The ratios of x(\hcfn)/x(\hcsn) are 3.6 on the average for all the starless cores and the Lup I-1 outflow source. While the other three outflow sources { show an average
 ratio of} 4.6. {
   For the ratio of x(\cts)/x(\hcsn), the starless cores and the outflow source Lup I-1 { have a ratio of} 0.4 on the average
   and the other three outflow sources 3.4, showing the largest difference among the compared ratios.
   These results indicate that the changes of the CCM emissions with the carbon length in Lup I starless cores and the outflow source Lup I-1 are different from those of the other three outflow sources. The three panels of Figure 4 clearly show that the correlations between each ratio of x(\hctn)/x(\hcfn), x(\hcfn)/x(\hcsn) and x(\cts)/x(\hcsn) are different for the starless cores/Lup I-1 outflow source and the other three outflow sources. The difference between the x(\cts)/x(\hcsn) of the Lup I starless cores/Lup I-1 outflow source and the other three outflow sources is the largest.}

\section{Discussion} \label{sec_discussions}

According to their different relative values of abundances of N-bearing CCMs and \cts,
the detected sources can be divided into three groups.
\begin{itemize}
\item[1] \textbf{Group CC} includes  all the five cold and quiescent dark cores in { the} Lup I region.
Their  abundances of N-bearing species decrease { with the length of the carbon chains}. The ratio  x(HC$_{2n+1}$N)/x(HC$_{2n+3}$N) (n=1-2) is 3.7 on the average.
Their ratio of { x(\cts~)/x(\hcsn~) is 0.4} on the average.
\item[2] \textbf{Group WC} contains Lup I-1 only, which is a known WCCC source. Its ratios of x(HC$_{2n+1}$N)/x(HC$_{2n+3}$N) (n=1-2)
are about the same as those of Group CC. However the ratios of { x(\cts)/x(\hcsn) is 0.3, larger} than those of Group CC, which means that \cts~
is more deficient in Lup I-1 than in Group CC sources.
\item[3] \textbf{Group JS} consists of I20582, L1221 and L1251A,  all with jets and shocks.
Their abundances of N-bearing species x(HC$_{2n+1}$N)
decline faster with the increase of `n' than the other two group sources.
Their { x(\cts~/x(\hcsn~)) ratio} is 3.4 on the average, which is in contrast
to the same ratios of the sources in the other two groups and it has not been seen previously.

Below we discuss physical conditions and chemical contents for each group.
\end{itemize}

     \subsection{Lupus I starless  dust cores}

    Lupus I is a nearby molecular cloud complex. 
       The distance measured by \citet{2008A&A...480..785L} is 155$\pm$8 pc. For our Lup I cores, except Lup I-1, all
       are dark and without associated stellar sources within radii of one arcmin except Lup I-5 which { contains} an object surveyed with Spitzer offset by $\sim$ 0.5 $\arcmin$ from the center \citep{2015A&A...584A..36G}  .
        The Herschel 250 \micron~images of Lup I cores are shown in the bottom panel of Figure A1.
       The dust temperatures of these cores range from 10.0 to 11.9 K, with an average value of 11.0 K which is similar to that of the
       cyanopolyyne peak in TMC-1, 10.6 K (\autoref{table_SED}).

		The dust cores Lup I-2 and Lup I-5 are located in the gas
		core C6 where \hctn~J=3-2 and 10-9 were detected \citep{2012MNRAS.419..238B}.  Lup I-6 was identified as a TMC-1 like cloud and is located in
        the C3 core where \hctn~J=10-9 was detected \citep{2010ApJ...718L..49S,2012MNRAS.419..238B}. The gas core C3 also includes the dust core Lup I-7/8/9,
        which was detected by our TMRT observations too. Lup I-11 is in the gas core C8 where \hctn~J=3-2 and 10-9 were detected \citep{2012MNRAS.419..238B}. It
        is also starless with IRAS 15422-3414 { located 62} arcsec away.
		
\begin{figure*}
\centering
\includegraphics[width=150mm]{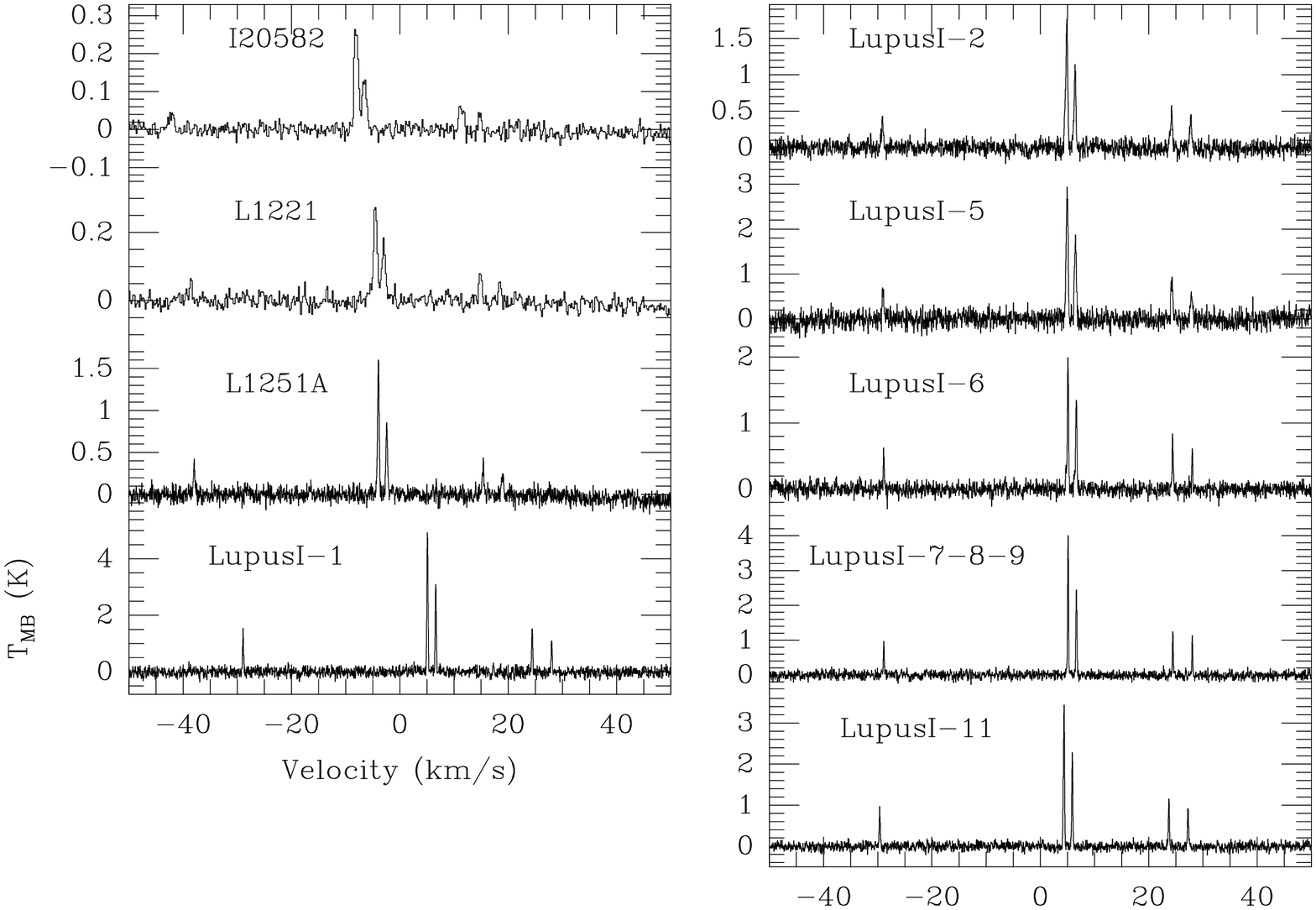}
\caption{\hctn~Spectra. \label{fig_hctn}}
\end{figure*}

From Figs. \ref{fig_hctn} and \ref{fig_spectra} and \autoref{table_obs_par}, one can see that CCM emissions of the four dust cores, Lup I-2, Lup I-5, Lup I-7/8/9
and Lup I-11 are all quite strong. Emissions of three (except Lup I-2) of them are even stronger than those of Lup I-6 (Lupus-IA) \citep{2010ApJ...718L..49S}.
These results demonstrate that these four starless dust cores detected with LABOCA, Herschel and Planck 
are all CCM emission cores.
The detections of fruitful CCM emissions in Lupus I starless cores as well as in
Lup I-1, the second WCCC source \citep{2009ApJ...697..769S},
indicate that Lupus I is a CCM rich region similar to Taurus.

There is no apparent ongoing star formation in these Lup I cores yet, except Lup I-1.
The chemistry of these cores in Lupus I  belong to that of a cold and quiescent phase.
In this early phase carbon atoms and ions needed to produce carbon-chain molecules arise from
photodissociation and photoionization during the diffuse phase of the
		cloud. { In such cloud} the C and C$^+$ from { a more diffuse earlier stage of evolution have not been incorporated into CO yet} \citep{1992ApJ...392..551S}.

{ All these starless cores and the outflow source Lup I-1 indicate { that} the Lup I is { a} rich CCM and popular carbon-chain--producing region, which { may match that of the Taurus molecular complex located at a similar distance from the Sun}. This region is located at the edge of the youngest subgroup (Upper-Scorpius) of the Scorpius-Centaurus OB association. It borders the expanding HI shell around the Upper-Scorpius at the north-east side which may represent the source of carbon atoms and ions to form CCMs in the Lup I region \citep{2015A&A...584A..36G,2009ApJS..185...98T,1992ApJ...392..551S}.}

{ The abundance ratios of N-bearing species x(HC$_{2n+1}$N)/x(HC$_{2n+3}$N) of Lup I starless cores are 3.8 and 3.5 (n=1,2) (Sect 3.4), which is not inconsistent with the values of 2-3 for n=1,2 in dark clouds \citep{1981A&A....99..239B}.


About the ratio of x(\cts)/x(\hcsn), the values of the Lup I starless cores 
 range from 0.3 to 0.5 with an average value of 0.4. It is noteworthy that the x(\cts)/x(\hcsn) ratio of Lup I-5 is the smallest, which is indicated by the blue point below the blue line in the right panel of Figure 4.}
The \cts~ emission of this core seems to be influenced by a { Spitzer detected} c2d source at the 0.5$\arcmin$ from the core center { \citep{2015A&A...584A..36G}}. 

{ Regarding the abundance ratio of { a source previously detected with the TMRT,} Serpens South 1a \citep{2016ApJ...824..136L}, the x(\hctn)/x(\hcfn) and x(\hcfn)/x(\hcsn) ratios are 6.5 and 2.0 respectively, and 4.3 on the average, which is close to those of the Lup I starless cores. The ratio of x(\cts)/x(\hcsn) is 1.0 which is larger than those of all the Lup I cores. Serpens South 1a is located in the Serpens South Cluster where young stellar objects are embedded. Infall motion was detected with \hcsn~ and NH$_3$. This may be related to the change of the x(\cts)/x(\hcsn) ratio.}

 \subsection{WCCC source Lup I-1}
         Lup I-1 is a WCCC source \citep{2009ApJ...697..769S} second only to 
 		 L1527 \citep{2008ApJ...672..371S}.
         WCCC sources contain stellar envelope regions with a slightly elevated temperature of $\sim$30 K. In this warmer environment,
 CH$_4$ can be evaporated from the dust grain
 		mantles into the surrounding gas, and reacts
 		with C$^+$ to produce carbon-chain molecules 
 		\citep{2008ApJ...681.1385H,2009ApJ...697..769S,1991MNRAS.249...69B,1984MNRAS.207..405M}.
  Lup I-1
       contains a young class 0 stellar object \citep{2015A&A...576A.109Y}.  The bolometric temperature is 52 K, higher than that of
       L1527 (44 K) \citep{2012A&A...542A...8K}.
       Besides the normal radiation of the protostar, a recent accretion burst likely happened during the last 10$^2$ to 10$^3$ yr. 
  This burst might lead to an increase of the stellar luminosity by a factor of 100, which would { heat the dust and enhance WCCC emission in this core \citep{2008ApJ...672..371S}. This was listed as one of the interesting perspectives raised by the recent episodic accretion burst by \citet{2013ApJ...779L..22J}}.

 Lup I-1 shows a young molecular outflow driven by { a} Class 0 object. The outflows traced with
 { CO} $J=3-2$ and $6-5$ { have an average} dynamical time $1.1\times10^3$ yr.
The dynamic time given by the lower-J CO line ($J=2-1$) is 2$\times$10$^3$ yr \citep{1996PASJ...48..489T}.
These time scales are all about one order of magnitude smaller than those of
 the WCCC source L1527 ($1.1\times10^4$ yr) \citep{2015A&A...576A.109Y}.

{ The x(\hctn)/x(\hcfn) and x(\hcfn)/x(\hcsn) ratios of Lup I-1 are similar to those of the Lup I starless cores. However the x(\hctn)/x(\hcfn) ratio of the median values { of 16} embedded low mass protostars is 8.7, and that of Orion is 13$\pm$6 \citep{2018ApJ...863...88L,1981A&A....99..239B}, showing that the changes of emissions from N-bearing species { with given length of the carbon chain differ} in different star formation regions.
In low-mass star formation regions,  the cyanopolyynes are usually quite deficient but still closely related \citep{2018ApJ...863...88L}.
The different x(\hctn)/x(\hcfn) { ratios} in starless cores and outflow sources may be caused by the different dominant pathways between \hctn~and \hcfn~\citep{1998A&A...329.1156T,2016ApJ...817..147T}.

The \cts~column density of Lup I-1 is lower than those of \hctn, \hcfn~and \hcsn, and the ratio of x(\cts)/x(\hcsn) is as small as that of Lup I-5, i. e. the smallest among our targets.}
         It shows that { the} \cts~abundance in Lup I-1 does not increase as much as in the other three outflow sources (see Sect. \ref{sec_JS_Chem})
though HH185 was found within the confines of the outflow region \citep{1996PASJ...48..489T}. 
The reason may be that the CCMs retained from the early core phase keep the abundances of these species at high levels, which might possibly happen during a fast collapse of the cloud's core like in Lup I-1 \citep{2008ApJ...672..371S}.
Another reason may be that the outflow is very young and there is not enough time for the formation of S-bearing LCCMs
from shock induced S$^+$.


	\subsection{Excitation conditions and chemistry in Group JS} \label{sec_JS}

	\subsubsection{{ Group JS}} \label{sec_JS_Phy}

		Three molecular outflow sources, I20582, L1221 and L1251A are included in the Group JS.

        In I20582,  molecular outflows
		were detected in the $J=1-0$ lines of CO and $^{13}$CO \citep{1989A&A...223..287H, 1995ApJ...454..345B}.
       The outflow age is $\sim10^5$ yr \citep{1997ApJ...491..653T}.	
		HH 199B1 to HH 199B3 are at or near the CO blue lobe detected with the OVRO interferometer \citep{2004ApJ...612..342A,1995ApJ...454..345B}.
		HH 199R1-R5 are located in the north-east and are distributed up to 8\arcmin~away from I20582.
        There is a chain of 2.122 \um~H$_2$ jet knots through the center and aligned in the south-east to north-west direction \citep{2004ApJ...612..342A,1995ApJ...454..345B}.
        The enriched infrared and optical characteristics indicate strong shocks in this source.
		The observation by \citet{1995ApJ...454..345B} revealed strong S$^+$ emission in HH 199B1-B3, B6 and HH 199R2.
        The peaks of the dense gas detected with CS J=2-1 coincide with the HH 199B1-B3 objects and the CO blue lobe peak, where there is a lack
        of quiescent gas traced by c-C$_3$H$_2$ \citep{1997ApJ...491..653T}.
	
        L1221 has the highest dust temperature $15.1\pm0.2$ K among
		all the LCCM detected sources.
	 CO $J=1-0$ outflows with an age $\sim 5 \times 10^4$ yr were detected with  the 45-m of Nobeyama Radio Observatory (NRO) \citep{1991ApJ...377..510U}.
        It was showing a U-shape distribution resulting from
		the interaction between the outflow and the surrounding gas \citep{1991ApJ...377..510U}.
		This core therefore appears to be cometary
		and contains the infrared source IRAS 22266+6845. There is a close binary consisting of two infrared sources located in the east and west of the IRAS source and the eastern one seems to drive the outflow \citep{2005ApJ...632..964L}. The source was
        detected in K$^\prime$ band and has a south-eastern extension in the same direction as the eastern lobe of the outflow,
        showing shocked H$_2$ emission. HH 363 seen in H$\alpha$ and S$^+$
		is also associated with the outflow \citep{1997IAUS..182P..51A}.
		Recently, three IRS sources were revealed with Spitzer
		IRAC/MIPS data and ground based submillimeter emissions.
		IRS 1 and IRS 2 are Class I objects and IRS 3 is a Class 0
		stellar object. IRS 1 contains an arc and IRS 3 a jet. IRS 3 also exhibits 3-6 cm emission which indicates shock ionization due to the interaction between the jet from IRS 3 and the surrounding gas \citep{2009ApJ...702..340Y}.


Finally, in the outflow source L1251A, a collimated molecular outflow was detected with CO $J=2-1$ line observation by
		\citet{2010ApJ...709L..74L}.
       The dynamical timescale of this outflow is $5.2\times 10^4$ yr assuming
        an inclination angle of 70$^\circ$ \citep{2010ApJ...709L..74L}. Spitzer IRAC
 		observations revealed four infrared sources IRS 1-IRS 4.
 		The driving source is IRS 3, a Class 0 object. From this object a
 		collimated long infrared jet originated. 
		Our observed position is 12 arcsec east and 9 arcsec south to
		IRS 3 (see \autoref{beams}). The detected strong Spitzer IRAC observations revealed a jet with a bipolar structure
        extending from north to south though IRS 3. The jet may originate from a paraboloidal
		shock \citep{2010ApJ...709L..74L}.


 { About the N-bearing species in these outflows, the average values of x(\hctn)/x(\hcfn) and x(\hcfn)/x(\hcsn) are 6.8 and 4.6  respectively, which are both larger than those of the Lup I starless cores and the WCCC source Lup I-1,
showing that the abundance of N-bearing species decreases faster with carbon length in these outflows than in starless cores and WCCC sources.
The ratio of x(\hctn)/x(\hcfn) is somewhat lower than the median value (8.7) of 16 low mass protostars detected by \citet{2018ApJ...863...88L}. 

 \autoref {table_calculated_values} shows that in the starless cores, x(\cts) is two or more times lower than x(\hcsn).
 The x(\cts) in the WCCC source Lup I-1 is also three times less than x(\hcsn). While in the other three outflow sources the x(\cts) are 2-5 times of the x(\hcsn).
These results demonstrate that the x(\cts)/x(\hcsn) values for the three outflows are larger than those of starless cores and the WCCC source almost one order of magnitude higher than their average value.}

\subsubsection{Chemical mechanism of Group JS sources} \label{sec_JS_Phy}

The { abnormal high abundance of \cts~compared with those of nitrogen-bearing LCCMs in JS sources}
has not been seen before.
Previous studies revealed that the column densities of C$_2$S have good positive correlation with those of \hctn~and \hcfn~in quiescent dark cores \citep{1992ApJ...392..551S}.
The column densities of \cts~are 3-5 times lower than those of C$_2$S \citep {1992ApJ...392..551S,1990A&A...231..151F}.
High C$_2$S and \cts~abundances were found towards Taurus dark clouds, and there is still a good correlation between column densities of C$_2$S and \hcfn.
In star formation regions, S-bearing CCMs were at best only marginally detected \citep{1992ApJ...392..551S},
indicating that the results of our three outflow sources are unusual.
WCCC theory was a possible explanation for the large hydrocarbon abundances such as \cfh~and C$_6$H present in L1251A \citep{2011ApJ...730L..18C}.
However, WCCC can not explain the { abundances of} \cts~in the three outflow sources, since the relative intensities of N-bearing and S-bearing molecules in WCCC sources are close to those in early cold quiescent cores as listed in \autoref{table_calculated_values}.

In the observations of \citet{2011ApJ...730L..18C} towards L1251A, the \cts~column density is lower than that of \hcsn, which is contrary to our results.
Their target point locates at  R.A.(2000)=22:30:40.4, Dec.(2000)=75:13:46,  while ours (\autoref{table_source} and \autoref{beams}) is more than one arcmin closer to IRS3 and the jet \citep{2011ApJ...730L..18C,2010ApJ...709L..74L}.
{ The associated outflow may heat the surroundings and contribute in inhibiting the depletion of S-molecules \citep{2016A&A...593A..94F,2018MNRAS.478.5514V}.}
We speculate that the high abundances of \cts~are relevant to the jets and shocks in the three sources.

These aspects adequately demonstrate that sulfur ions are produced during shock processes
developed in these three sources. Shock processes may lead to the reduction of CCMs. However, shocks can fuel the environments with plenty of S$^+$ and thus driving the generation of S-bearing CCMs including \cts. The regions with most abundant S-bearing CCMs do not necessarily completely coincide with the shocked regions since sulfur elements will only enter into S-bearing CCMs through mild chemical reactions. Unlike in cold dark clouds and WCCC sources, shock induced chemistry plays a major role in performances of N- and S-bearing CCMs in these three { outflow} sources. This is a new chemistry and we name { it shocked} carbon-chain chemistry (SCCC). The characteristics of our SCCC sources are as following:

        1. Their emissions of the N-bearing species are generally weaker than those in early cold and dark cores and WCCC sources.
           Taking our samples including three SCCC sources and six of the early cold/WCCC sources into account only, the highest column density of \hctn~characterizing the SCCC sources  is close to the lowest value of the early cold/WCCC sources.

        2. They have relatively strong \cts~emission.
        Different from the early cold and dark cores and WCCC sources, the column densities of \cts~of SCCC sources exceed those of \hcsn, and this could be used as a criterion to identify SCCC source.

        3. The sources are associated with molecular outflows and infrared/optical jets. The dynamic time scales range from  5$\times$10$^4$ to 10$^5$ yr.

        4. Emissions of ionization species, especially S$^+$, are enhanced in shocked regions.



\subsection{Model test} \label{sec_JS_Chem}
We modeled the chemistry of WCCC/SCCC sources by treating them
as homogeneous, isotropic clouds.
In this simulation, a single point chemical network was run under an ordinary differential equation solver
DVODE \citep{doi:10.1137/0910062} with most physical parameters
fixed, n(H$_2$) = 10$^5$ cm$^{-3}$, A$_V$ = 5 mag,
$\sigma$$_g$ = 0.03 $\mu$, and cosmic-ray
ionization rate $\gamma$ = $1.2\times10^{-17}s^{-1}$ \citep{2004ApJ...617..360L}.  The metal abundances were adopted as the low-metal
abundance case of \citet{1982ApJS...48..321G}, except for that of sulfur.
In star-forming regions, the total abundance of sulfur-bearing species is $\sim$ $10^{-8}$ \citep{ 2016ApJ...824..136L},
and this value was adopted instead of the widely used $8\times 10^{-8}$ \citep{1998A&A...334.1047L}.
All the elements were initially ionized except for the hydrogen atoms.
The network of chemical reactions as well as the binding energies of the species were adopted from the UMIST
Database for Astrochemistry (http://www.udfa.net) as described in \citet {2013A&A...550A..36M}.
The network consists of 6173 gas phase reactions and 195 reactions on grain
        surfaces involving 467 atomic and molecular species.
The accretion and desorption rates were calculated according to \citet{ 1992ApJS...82..167H}, assuming a sticking coefficient of { unity}.
The temperatures of the gas and dust were assumed to be well coupled.
The changes of the fractional abundances of HC$_{2n+1}$N (n=1-3) and \cts~with time are shown in the left panel of \autoref{fig_model}.


The temperature, represented by the dotted green line in the left panel of \autoref{fig_model}, stays as 15 K till $10^5$ yr after the beginning of modeling.
Apparently
abundances of N- and S-bearing CCMs would increase first and then  get depleted significantly before $10^5$ yr. During this period,
the abundance of \cts~is positively correlated with those of N-bearing CCMs. These characteristics agree with the observations in dark clouds and
star-forming sources \citep{1992ApJ...392..551S}.

As the temperature increases to 50 K at $1\times 10^5$ yr, abundances of N-bearing species increase quickly because of the reactions between the durable nitrogen atoms and the evaporated molecules such as CH$_\mathrm{4}$ and $\mathrm{C_2H_2}$.
Soon after, abundances of N-bearing CCMs drop slowly for the lack of sustained supplements of precursors.
The enhancements of N-bearing CCMs with shorter carbon { chains}, especially \hctn, are more significant when compared with heavier N-bearing LCCMs.
Such { a} behavior can explain why WCCC sources such as L1527 and our Group JS sources have relatively large x(HC$_{2n+1}$N)/x(HC$_{2n+3}$N).
However, the abundance of \cts~is not enhanced much during this period because sulfur atoms have { not been
consumed yet, are not as effectively incorporated} as nitrogen atoms,
and  the downward trend of \cts~can not be reversed.
These results indicate that, besides hydrocarbons such as \cfh~detected in WCCC sources,
the N-bearing CCMs will also be largely enhanced in the early phase of WCCC.  It can explain the variances between the younger WCCC source Lup I-1 \citep{2009ApJ...697..769S} and the older one L1527 \citep{2008ApJ...672..371S}  whose
abundances of N-bearing CCMs are nearly an order of magnitude lower than those of the former one.

Shocks can be the driving sources for a persistent supply of sulfur atoms and ions.
The right panel of \autoref{fig_model} shows the { evolution} of the abundances of some important species after a J-shock simulated by the MHD shock code of \citet{2015A&A...578A..63F}.
The simulation ran with a  magnetic field parameter b = 0.1, a pre-shock density n$_H$ = 10$^5$ cm$^{-3}$, and shock speed, $v_s$ = 5 km\ s$^{-1}$ as described in \citet{ 1992ApJS...82..167H}.
The abundances of $N^+$, $C^+$ and S$^+$ would peak at  5$\times$$10^2$ yr after the shock, and then decrease because of the piling up of the after-shock fluid and contribute in inhibiting the depletion rates.
In realistic cases in molecular clouds, after-shock material can be spread diffusely and bring the shock-induced ions into the environment.
A simple order of magnitude estimation can be made to support the idea that the shock can provide enough S$^+$ to drive the evolution of sulfur-bearing carbon-chain molecules.
Supersonic shocks with a velocity $v_{e}$ = 5 km s$^{-1}$ are powerful enough to spur sulfur elements from grain surfaces and
to  release shocked gas into the environment with S$^+$ at abundance $f_s=10^{-9}$ (the right panel of \autoref{fig_model}). For a jet with a mass flux ($S_s$) $10^{-7}$ $M_{\sun}$/yr  and
a velocity ($v_s$) 300 km s$^{-1}$, the rate of the supplement of sulfur ions to the environment can be estimated as
\begin{equation}  f_s(\frac{v_s}{v_e})^2\frac{S_s}{M_{env}}\ \sim\ 10^{-20}\ s^{-1}\ per\ H\ atom \end{equation}
in which the envelope mass, $M_{env}$, is taken as 1 solar mass.

After $2\times 10^5$ yr, a supplement of S$^+$ is applied at a rate of $10^{-20}$ $s^{-1}$ per H atom.
The abundance of \cts~rises by one order of { magnitude and quickly exceeds} that of \hcsn.
The artificial supplement of S$^+$ is critical for explaining the emission { characteristics} of S- and N-bearing CCMs in SCCC sources,
since WCCC alone can not reproduce the high abundance of \cts.
The abundance of \cts~even exceeds the emission of \hcfn~at 2.5$\times$10$^5$ yr if the duration of supplementing of S$^+$ is not limited.
The source L1251A is close to that point with a \cts~column density comparable to that of \hcfn.
The time duration between this point and the start of the SCCC process  (5$\times$10$^4$ yr)  is comparable to the dynamical timescale of L1251A  (5.2$\times$10$^4$ yr; Sect \ref{sec_JS_Phy}).
The column density of \cts~may exceed that of \hcfn~in more evolved SCCC sources.

 Shocks are more effective in supplying S$^+$ than C$^+$ and N$^+$ (see the right panel of \autoref{fig_model}).
Our sources in { the} JS group are all associated with HH objects and jets harboring { H$\alpha$ lines and \ion{S}{II} features in emission}, and shocked regions are reservoirs of $S^+$ \citep{1995ApJ...454..345B,1997IAUS..182P..51A,2010ApJ...709L..74L}.
The precursors of N-bearing CCMs such as N and C$^+$ are more abundant compared with S$^+$
(see the left panel of \autoref{fig_model}) which makes the shock induced enhancements of N-bearing LCCMs
relatively ineffective.
On the other hand, the pre-consuming of precursors such as CH$_4$ in { the} WCCC process would further restrain the enhancements of N-bearing LCCMs.
The abundances of N-nearing CCMs are nearly unaffected if
supplement rates of C$^+$ ($10^{-21}$ s$^{-1}$ per H atom) and N$^+$ ($10^{-22}$ s$^{-1}$ per H atom)
are added in the gas-grain model after $2\times 10^5$ yr.
These characteristics imply that sulfur-containing species are unique tracers of shocks.

\section{Summary}\label{sec_summary}
	
	Using the new TMRT telescope of the Shanghai Observatory, 11 sources including five
	 outflow sources and six Lupus I starless dust cores were observed measuring
	\hctn~$J=2-1$, \hcfn~$J=6-5$, \hcsn~$J=14-13,~15-14,~16-15$ and \cts~$J=3-2$.
    Four of the five outflow sources, I20582, L1221, L1251A and Lup I-1 were detected.
 Among them, I20582 (IRAS20582+7724) and L1221 were newly detected as carbon-chain--producing regions.
    In the six starless cores of the Lupus I region all the observed transitions were detected except Lup I-3/4.
    For our detected sources, the excitation temperatures derived from hyperfine component fitting of \hctn~$J=2-1$ are consistent with the dust temperatures derived from Herschel data.
  Column densities are calculated.  Abundances were derived and analyzed for different kinds of species and sources. The main results are as { follows}:

1. Emission of \cts~$J=3-2$ in the three outflow sources I20582, L1221 and L1521A was found stronger than their \hcsn~emissions, which can not be explained by chemistry of early cold and dark cores and warm carbon-chain chemistry (WCCC) sources. Shock carbon-chain chemistry (SCCC) is suggested.
   In SCCC sources, shocks fuel the environments with abundant S$^+$ and thus drive the generation of S-bearing CCMs including \cts.

2. In the other outflow source Lup I-1 all the observed lines are strong.
However, emission of the \cts~is still weaker than those of the detected N-bearing species in this source, similar to the cases in the starless cores in the Lupus region. 
In young SCCC sources such as L1251A, WCCC can still play a role in local regions.

3. Emissions of N- and S-bearing CCMs of starless cores in Lupus I are pretty strong, and those of Lup I-11, Lup I-5 and Lup I-7/8/9 are even stronger than those of the TMC-1 like cloud Lup I-6 (Lupus-1A).
The results of these cores and Lup I-1 show that Lupus I is a carbon-chain molecule({ CCM}) rich complex similar to the Taurus complex.

4. Our gas-grain chemical model shows that SCCC is necessary to explain the observations toward the three above mentioned outflow sources since
   the WCCC can enhance the abundances of N-bearing CCMs while it does not strongly influence the S-bearing CCMs.
   Shock feedback by protostars can provide enough S$^+$ to drive SCCC.
    Our results demonstrate that the chemistry of S- and N-bearing species can be different in molecular outflow/jet sources.
More observations are needed to further explore SCCC.

\section*{Acknowledgements}
	
	We are grateful to the staff of SHAO and PMO Qinghai Station. We also thank Shonghua
	Li, Kai Yang and Bingru Wang for their assistance during
    observation period. This project was supported by the grants
	of the National Key R\&D Program of China No. 2017YFA0402600,
    NSFC Nos. 11433008, 11373009, 11503035, 11573036 and U1631237, and the China Ministry
	of Science and Technology under State Key Development Program
	for Basic Research (No.2012CB821800), and the
	Top Talents Program of Yunnan Province (2015HA030). D.Madones acknowledges support from CONICYT project Basal AFB-170002.
    Natalia Inostroza acknowledges CONICYT/PCI/REDI170243.

\bibliographystyle{mnras}

\bibliography{ms}	

\begin{figure*}
\centering
\includegraphics[width=0.30\textwidth]{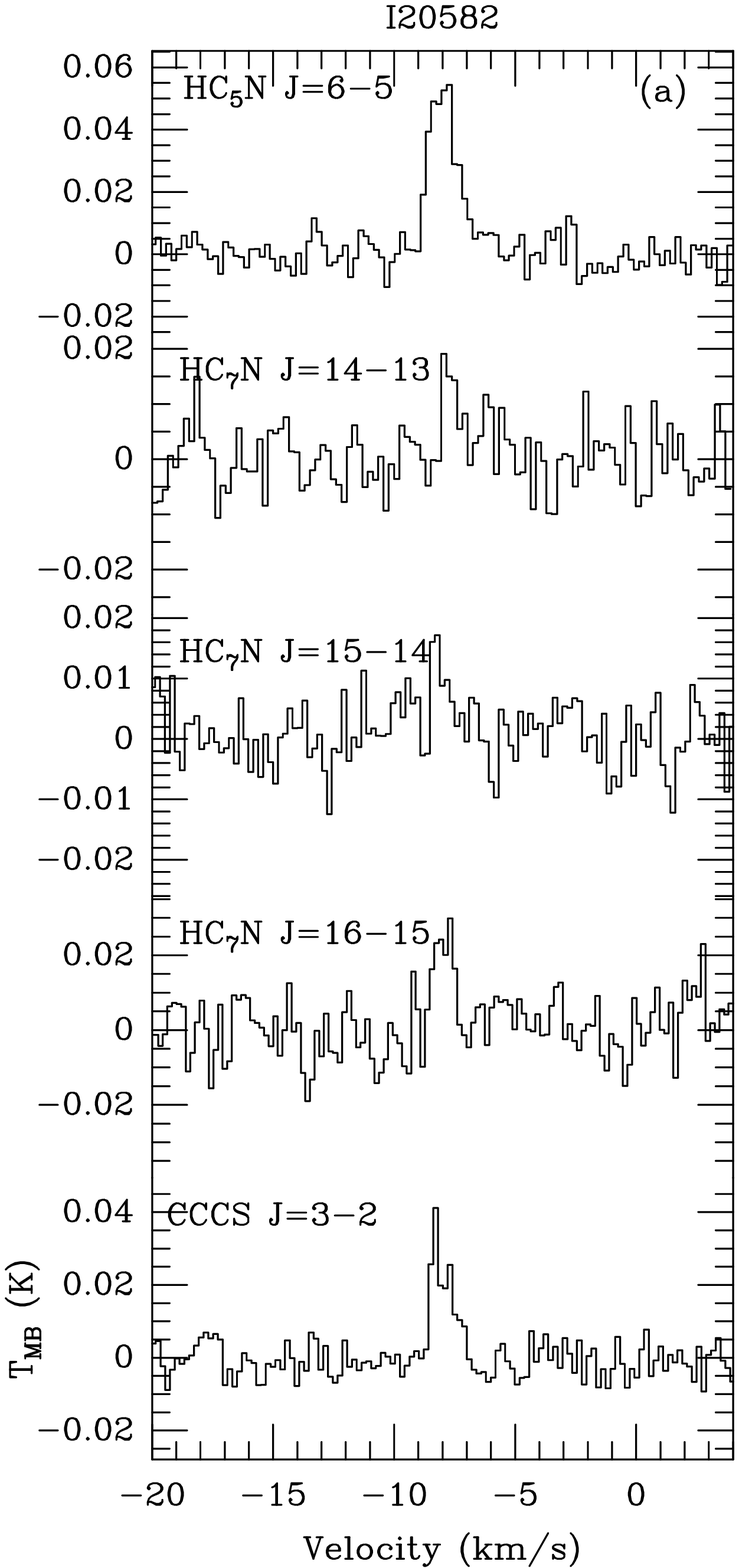}
\includegraphics[width=0.30\textwidth]{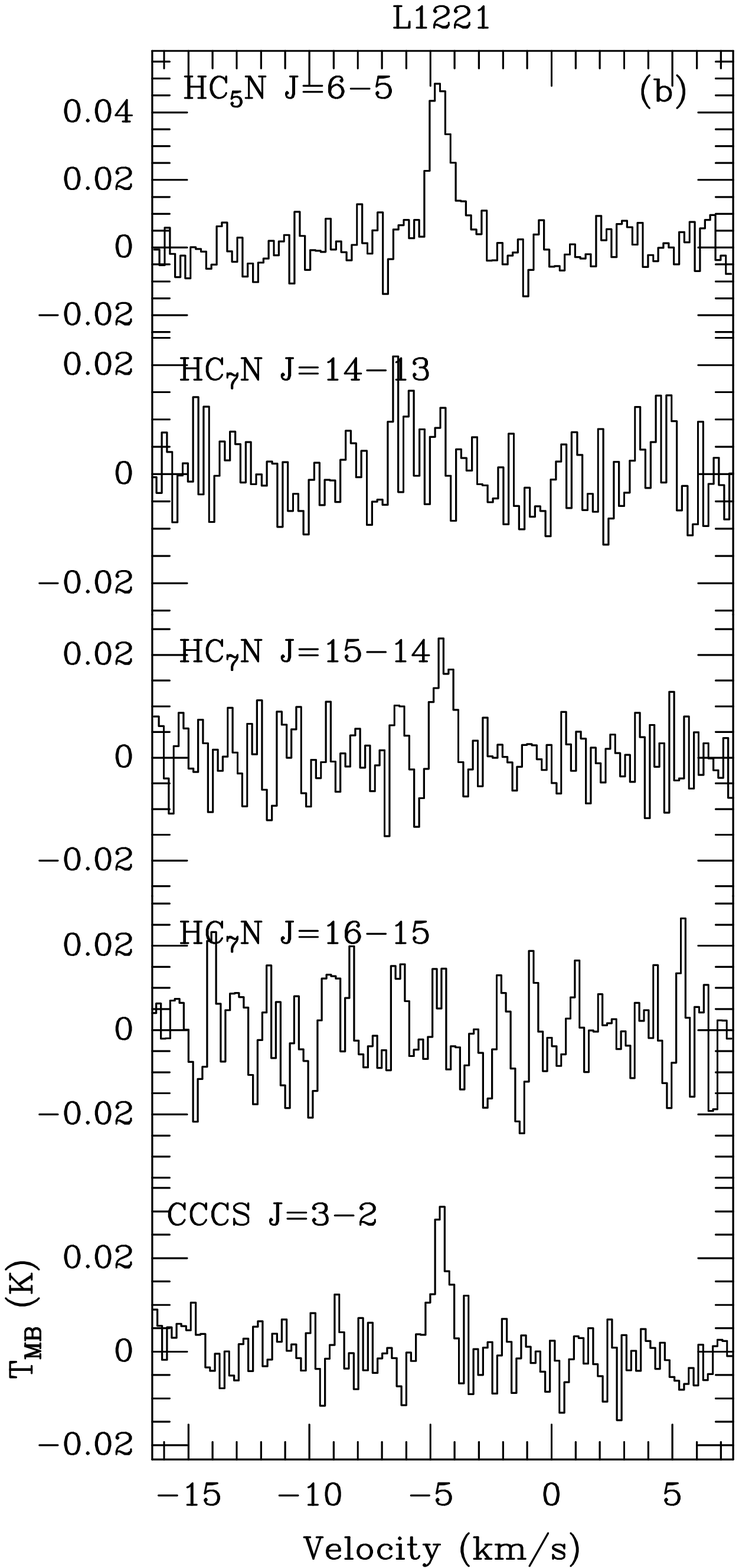}
\includegraphics[width=0.30\textwidth]{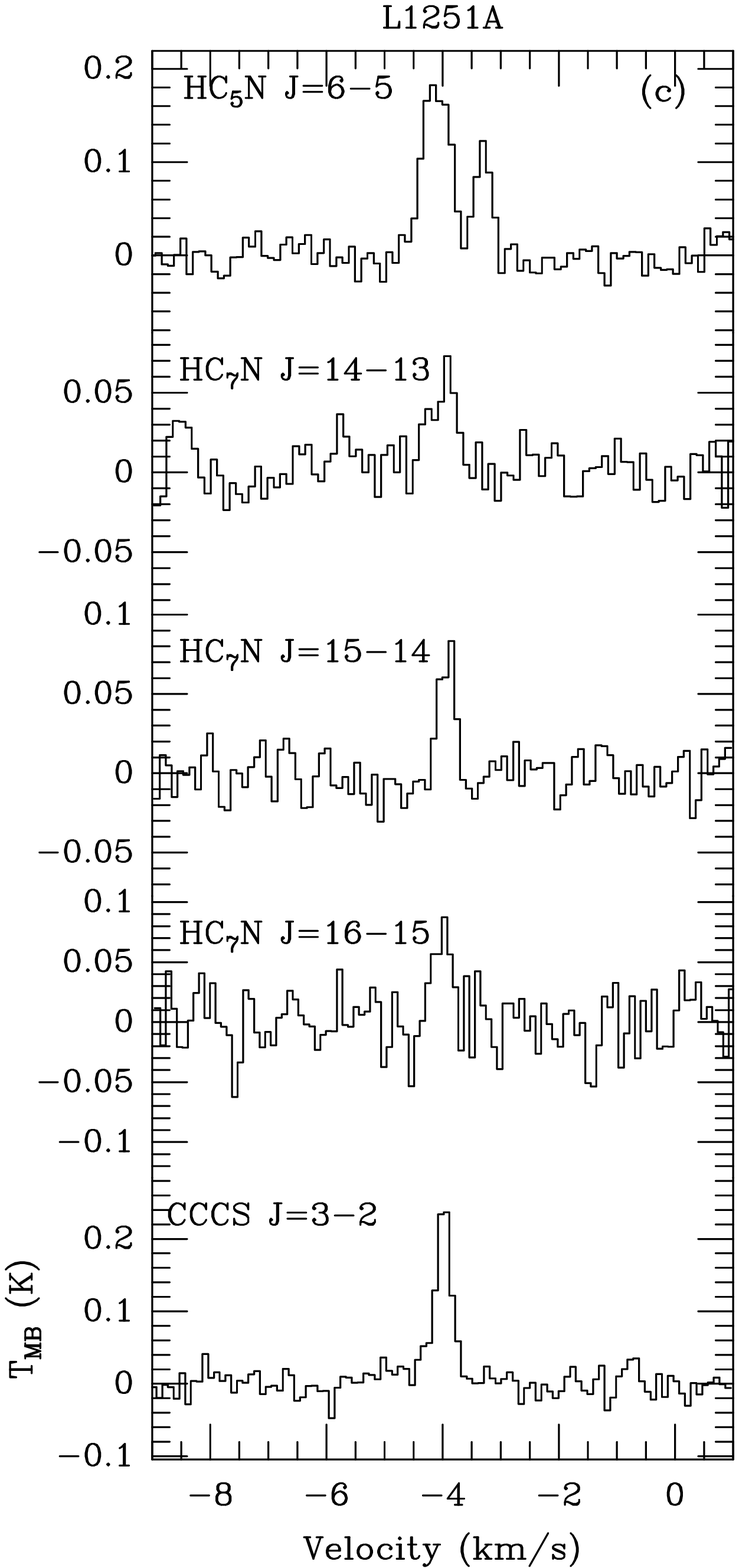}
\includegraphics[width=0.30\textwidth]{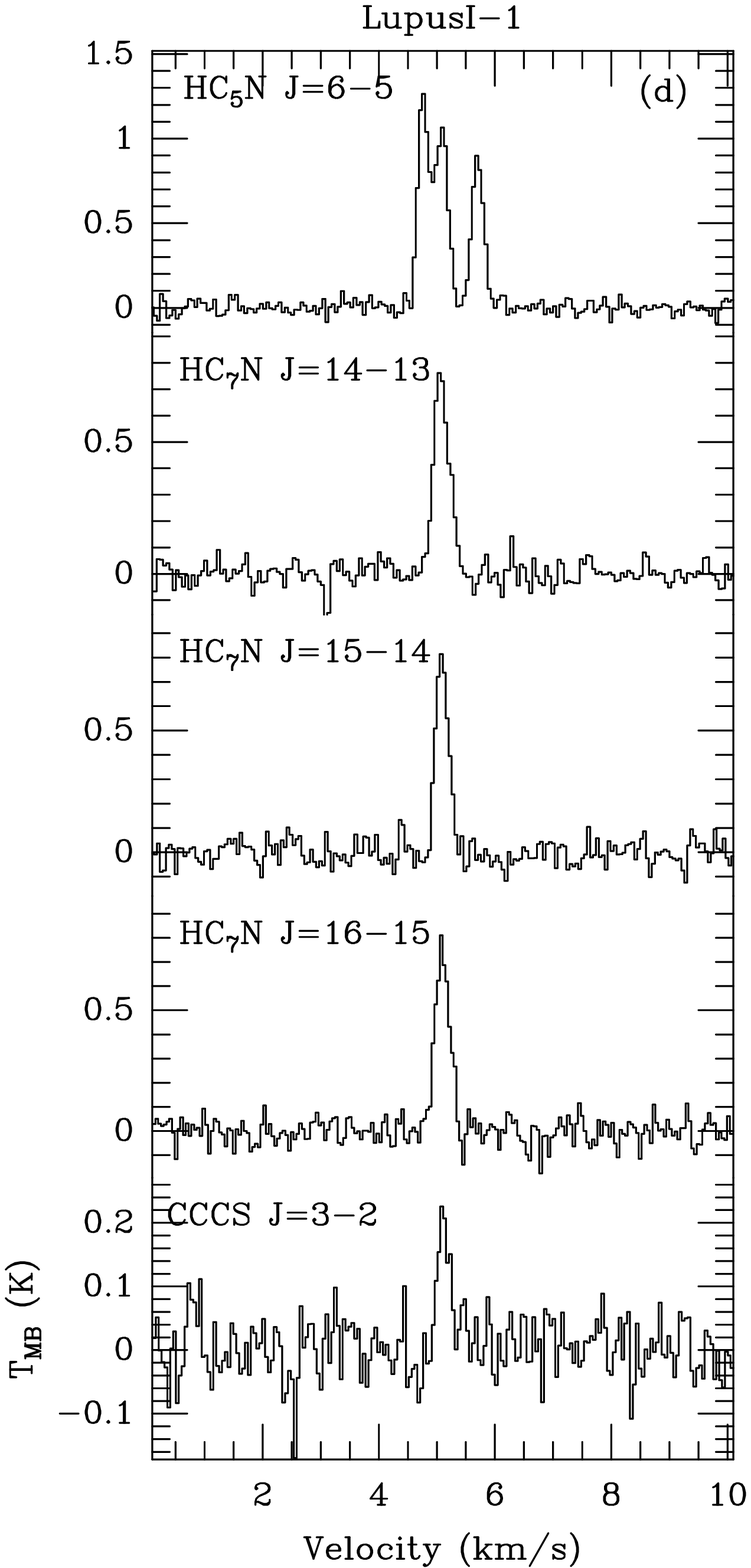}
\includegraphics[width=0.30\textwidth]{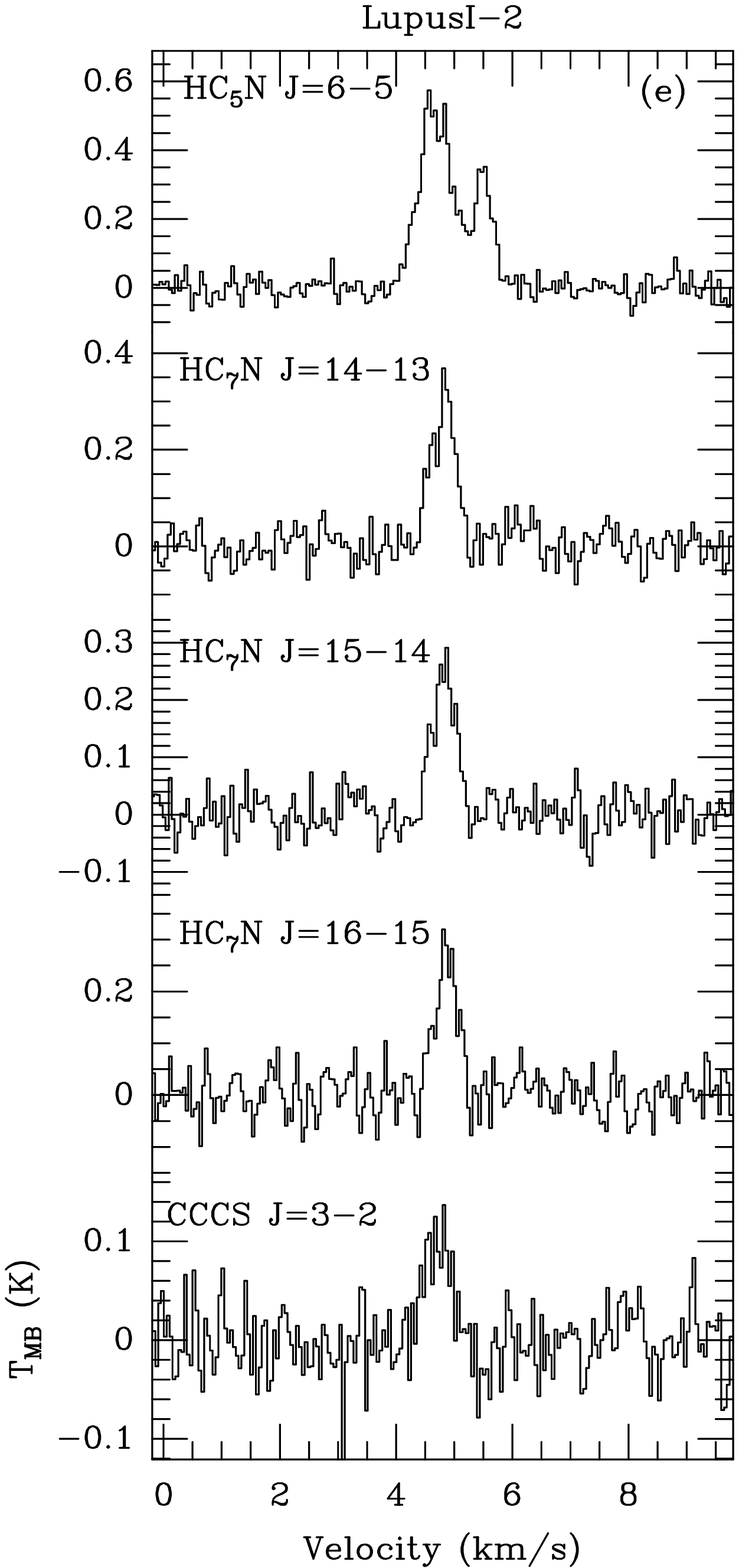}
\includegraphics[width=0.30\textwidth]{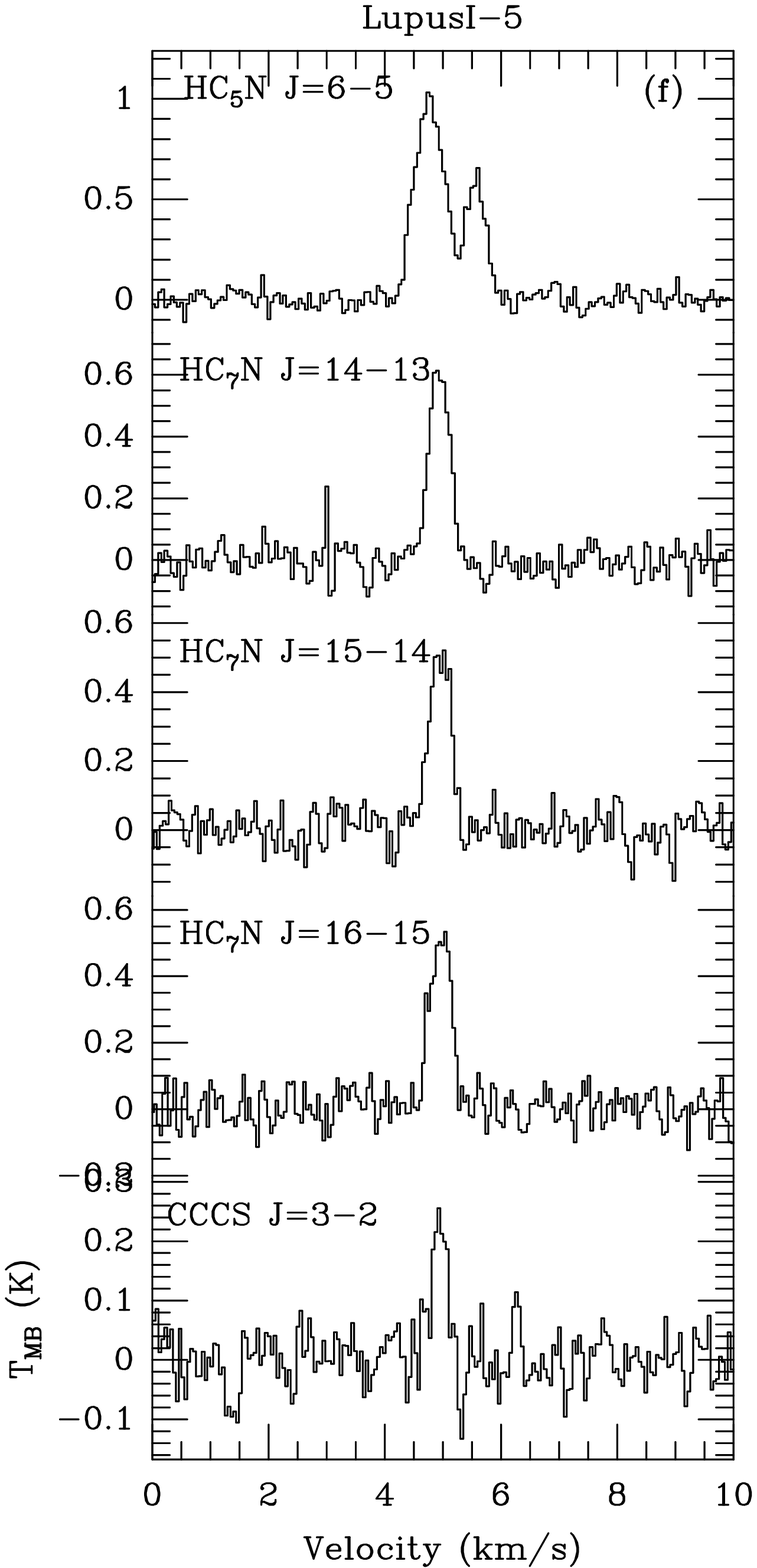}
\caption{Spectra of \hcfn, \hcsn~and \cts. \label{fig_spectra}}
\end{figure*}

\addtocounter{figure}{-1}

\begin{figure*}
\centering
\includegraphics[width=0.30\textwidth]{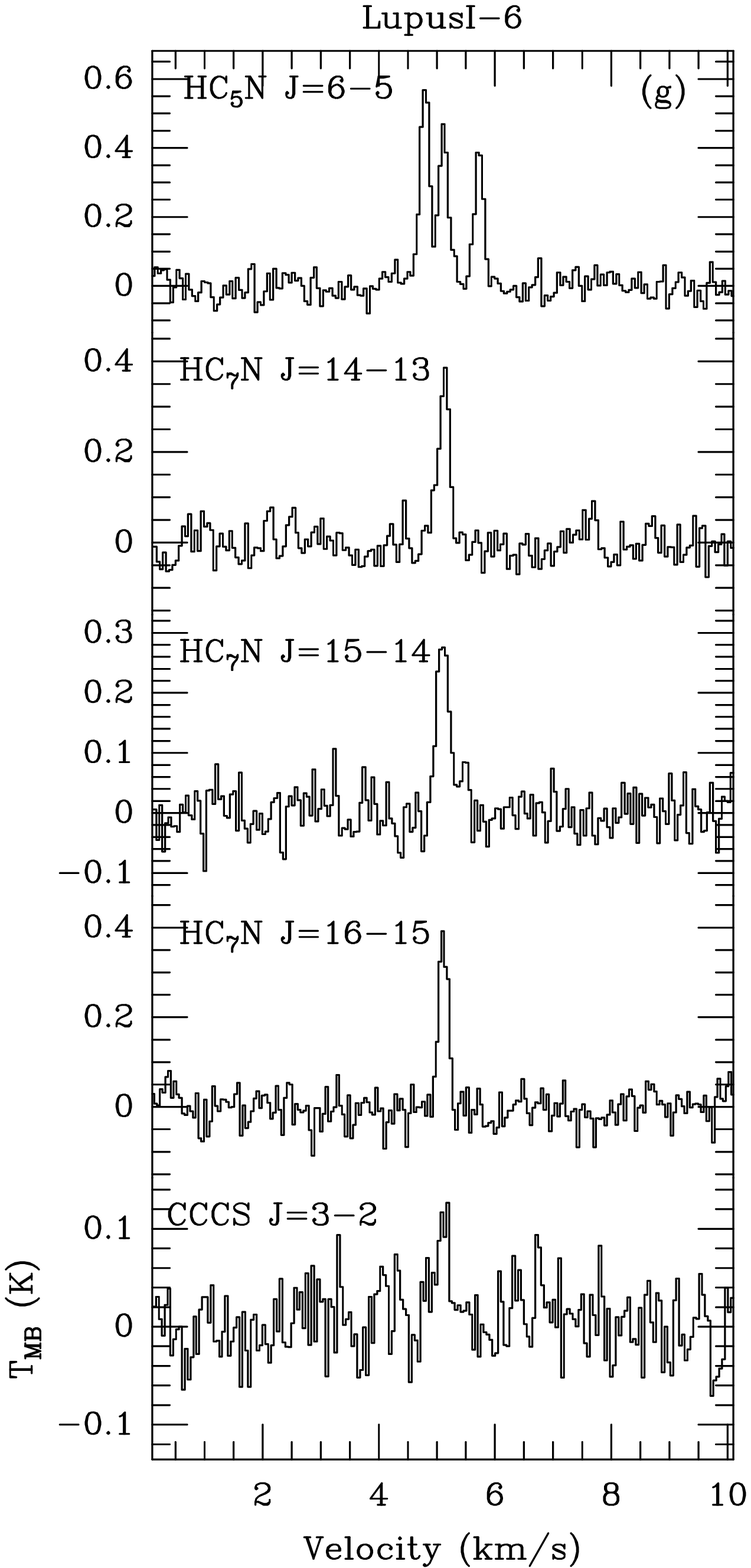}
\includegraphics[width=0.30\textwidth]{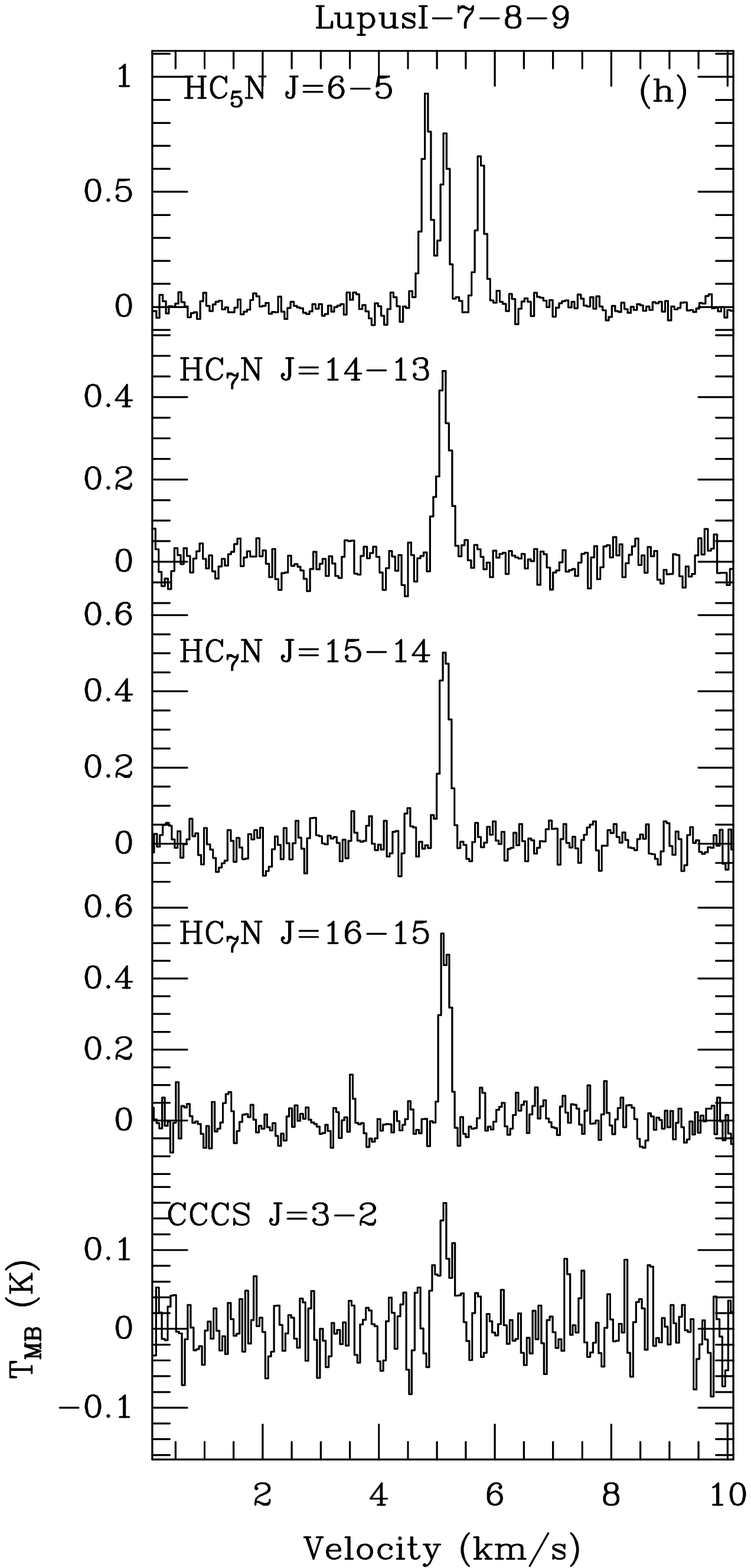}
\includegraphics[width=0.30\textwidth]{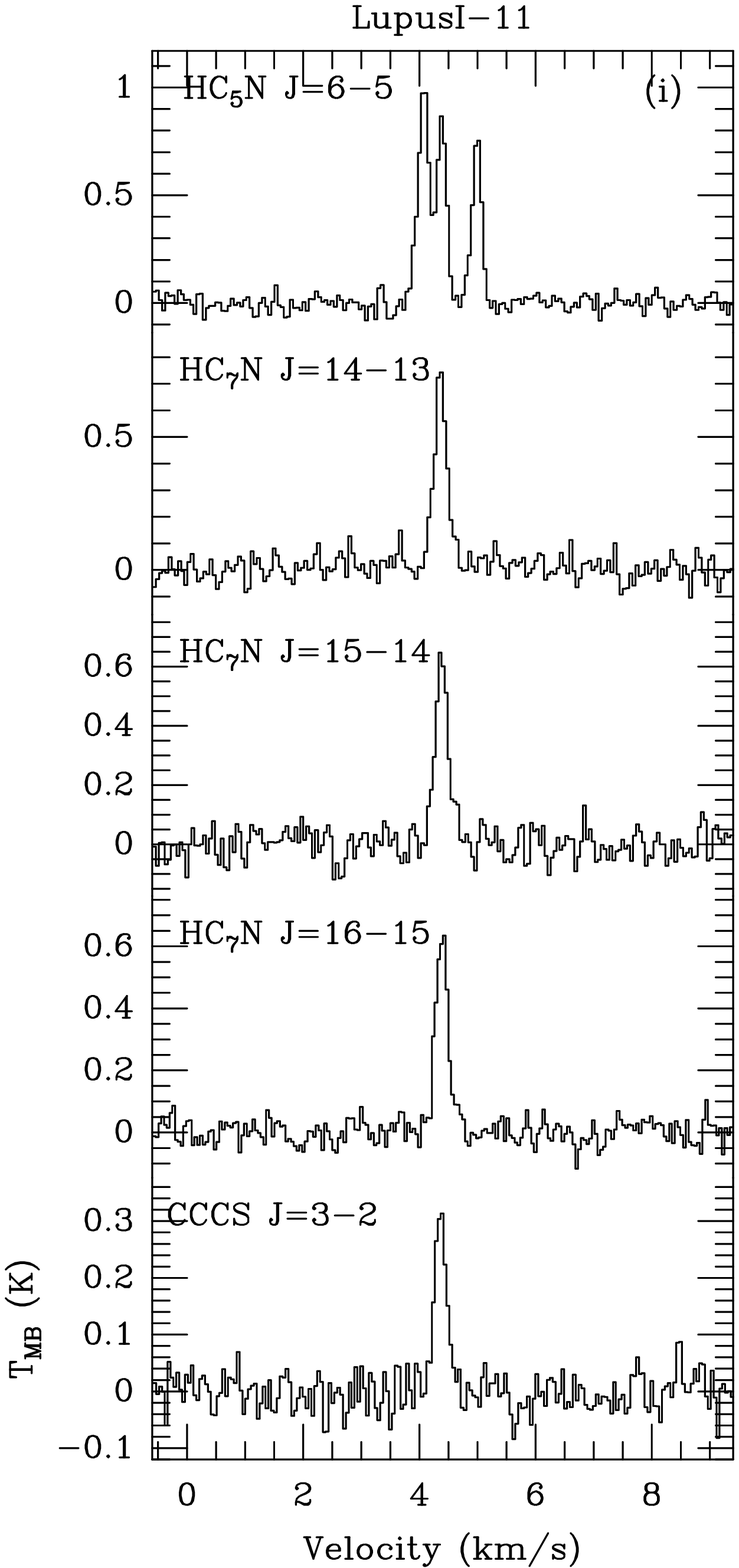}
\caption{Continued.}
\end{figure*}

 \begin{figure*}
     	\centering
        \includegraphics[width=0.48\textwidth,height=0.68\textwidth]{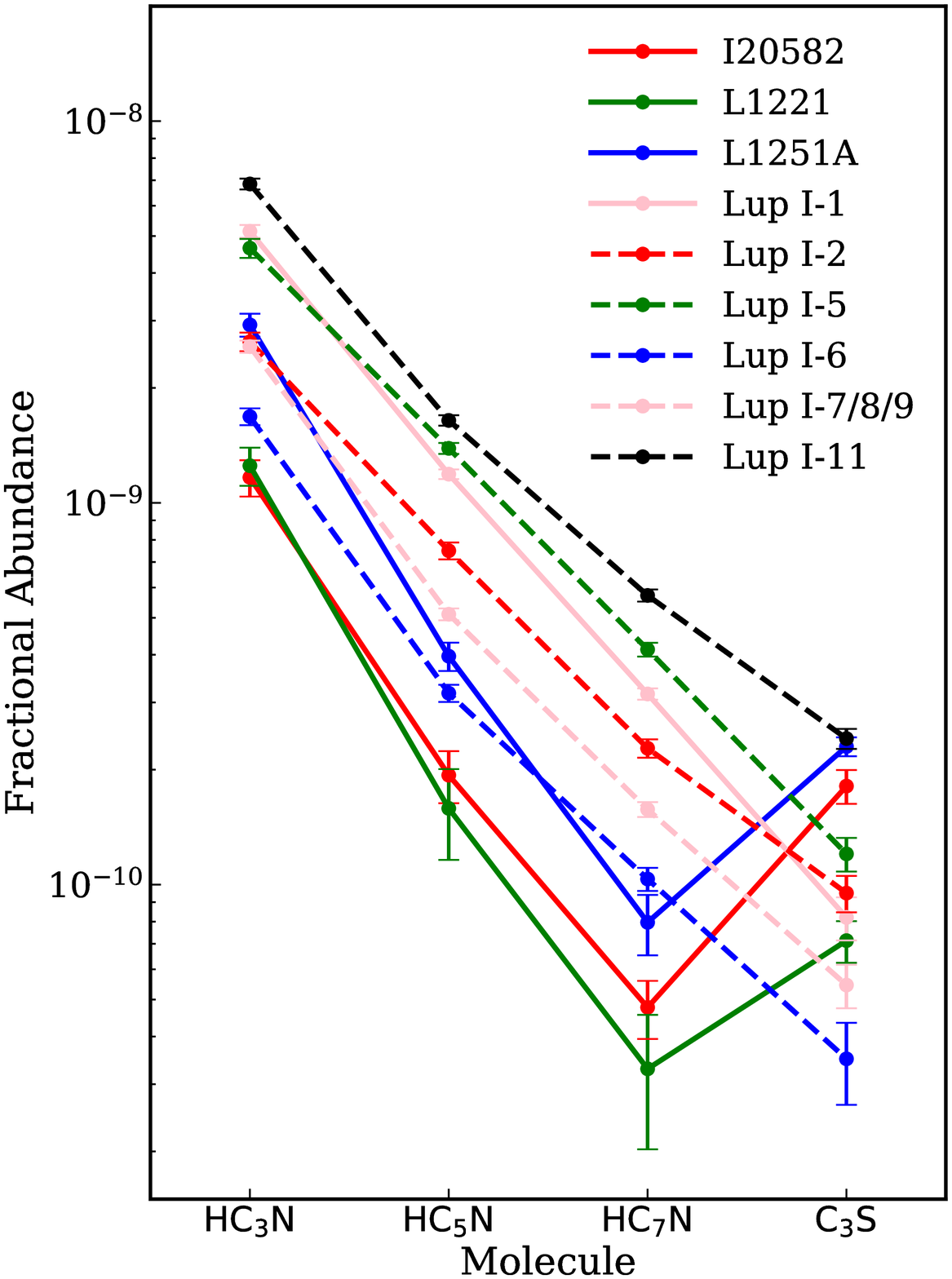}
     	\caption{ 
                  Abundances of detected species.
                  \label{fig_abundance} }
\end{figure*}

\hspace {3cm}

\begin{figure*}
\centering
\includegraphics[width=0.30\textwidth]{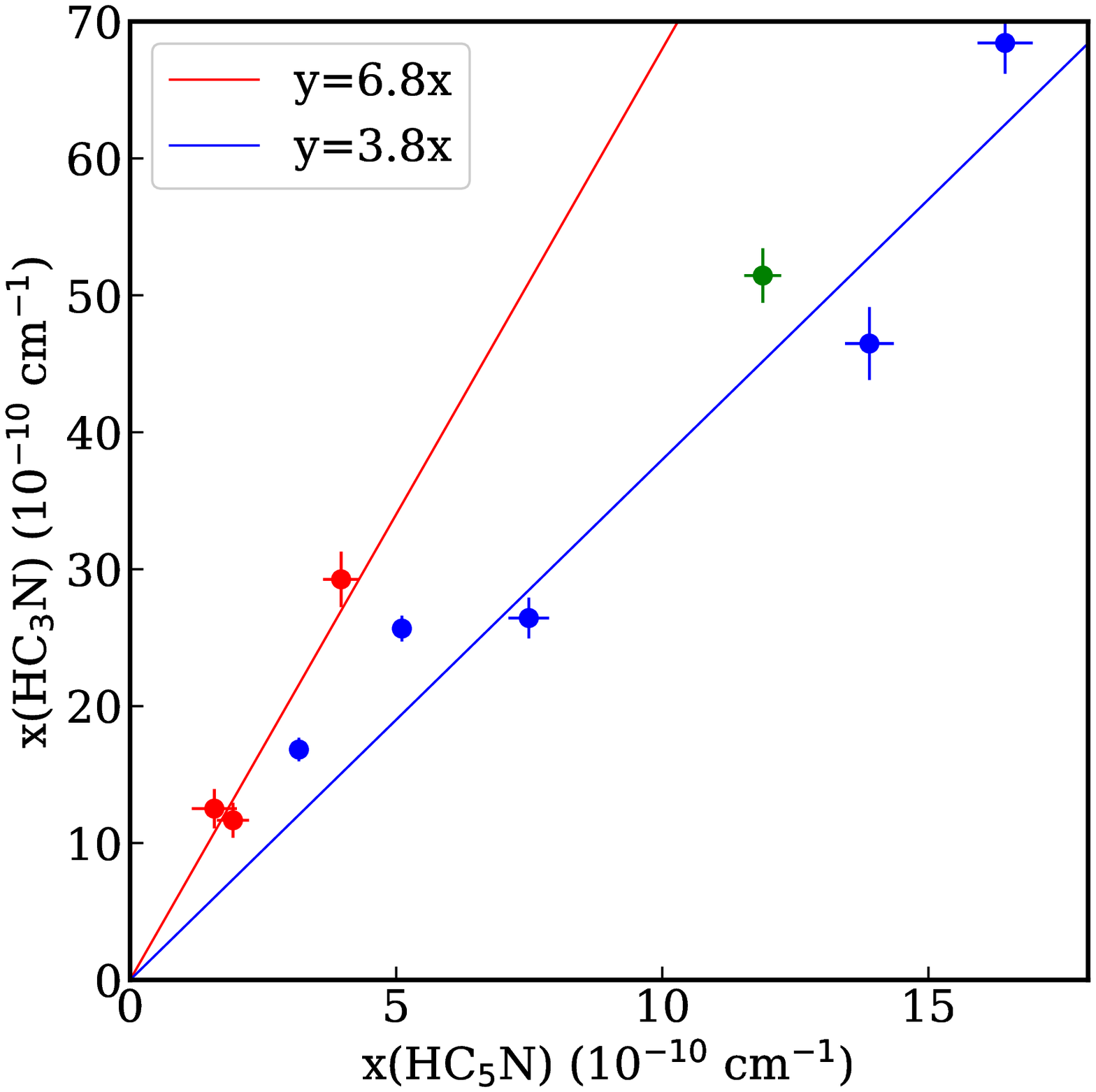}
\includegraphics[width=0.30\textwidth]{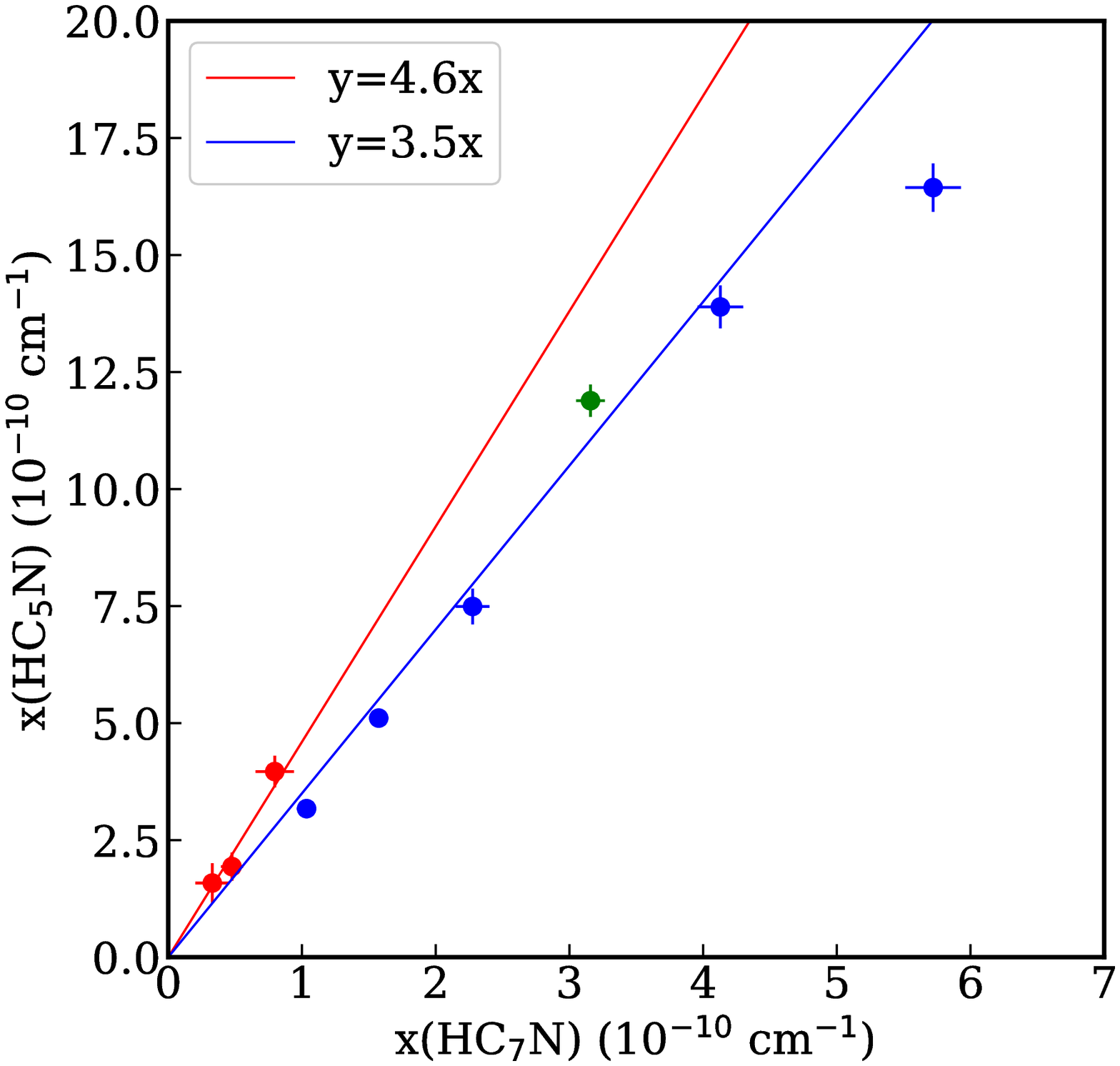}
\includegraphics[width=0.30\textwidth]{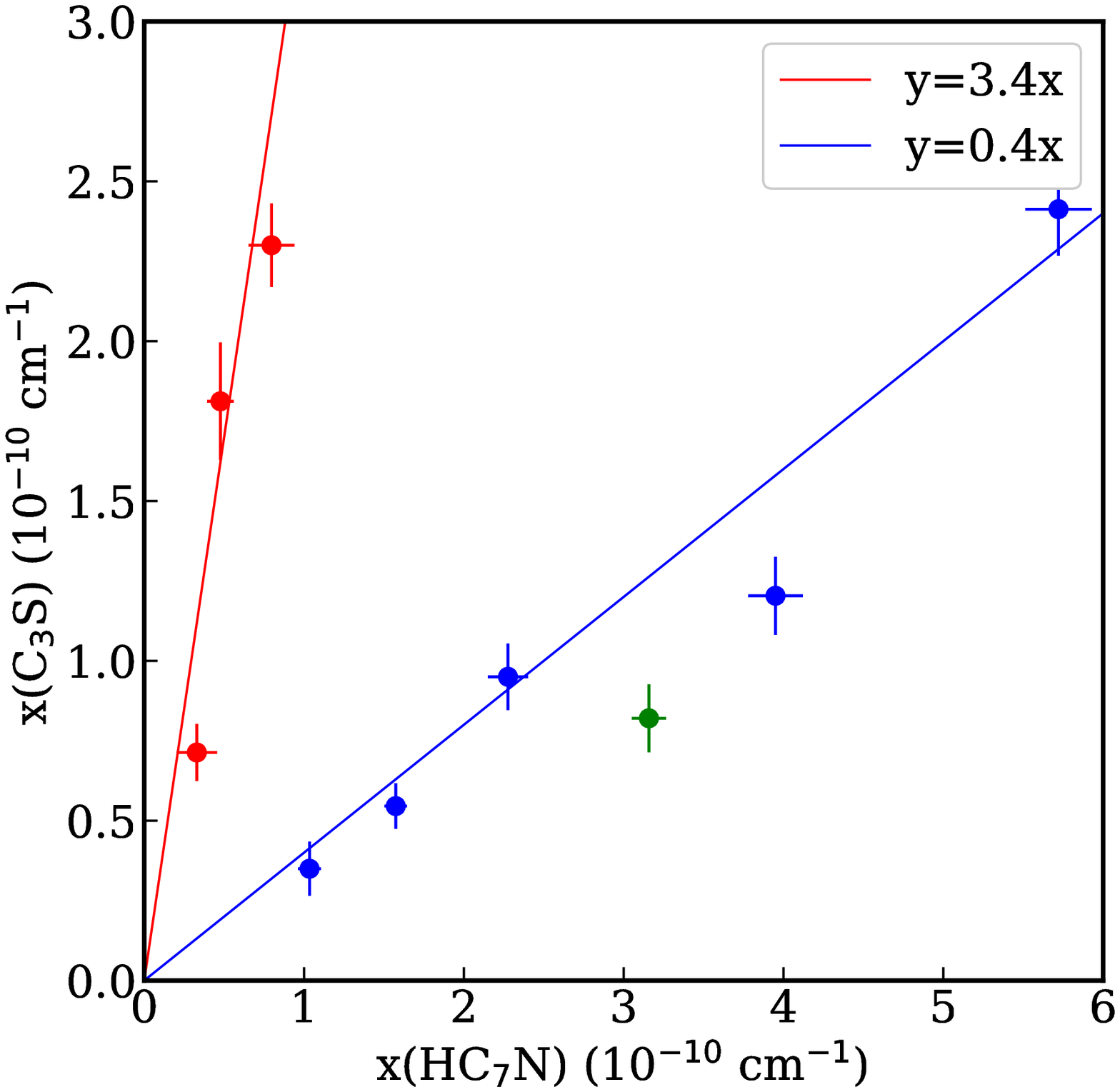}
\caption{Abundance ratios of different species. { \label{compare}
Left: x(\hctn) to x(\hcfn); Middle: x(\hcfn) to x(\hcsn); Right:
 x(\cts) to x(\hcsn)). For each panel, the blue line shows the result of linear fitting in bi-logarithm coordinates,
  and the red line indicates y=$<\frac{y_i}{x_i}>$x, where $x_i$ and $y_i$ denote the corresponding abundances for plotted dots.
In all panels the blue dots label all the Lup I starless cores, the green dot shows Lup I-1. In this figure the red dotes denote the other three outflows, which were excluded by the correlation between the two parameters of all the Lup I sources.}}
\end{figure*}

\clearpage

\begin{figure*}
     	\centering
         \includegraphics[width=0.49\textwidth]{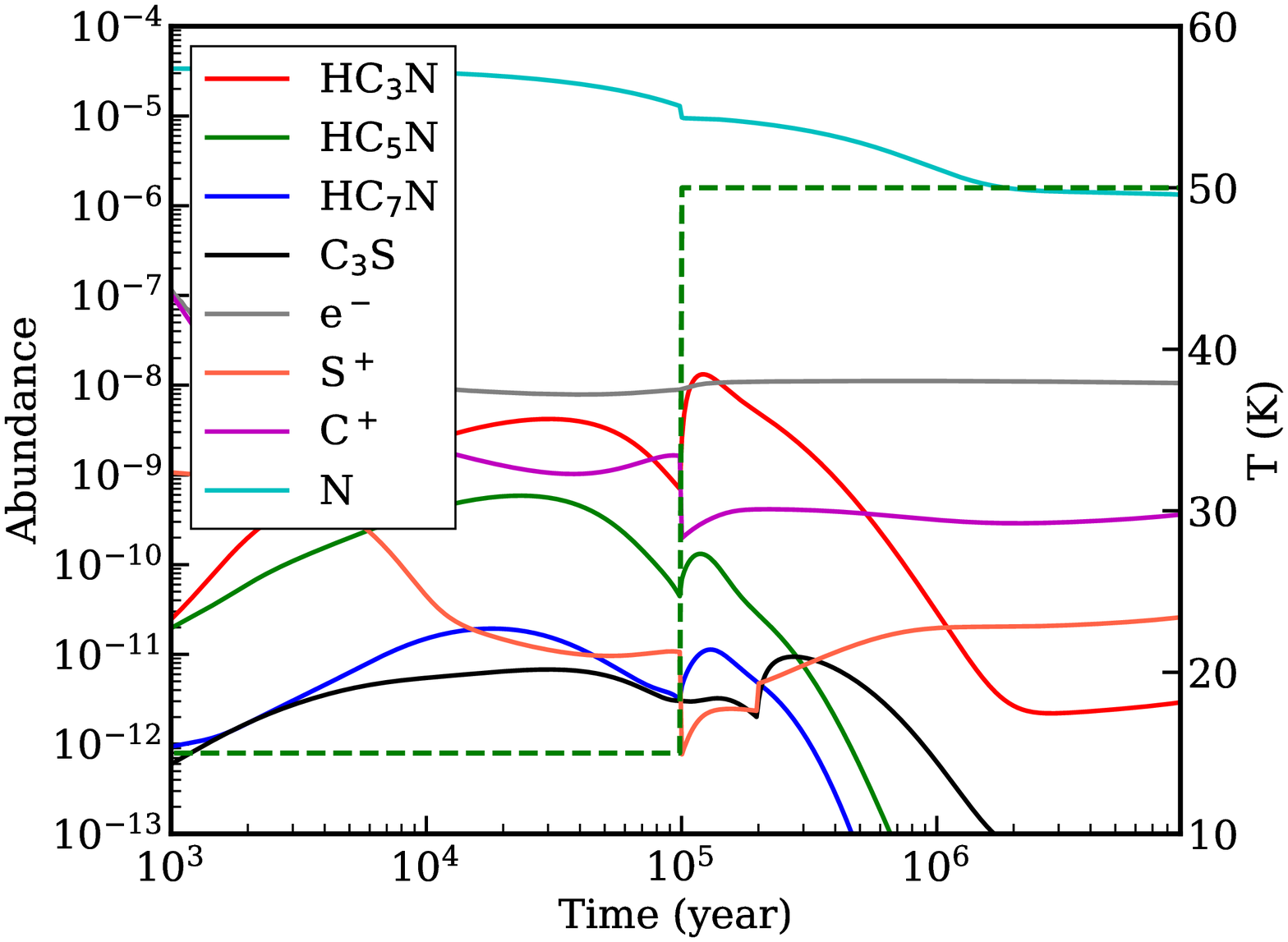}
                  \includegraphics[width=0.49\textwidth]{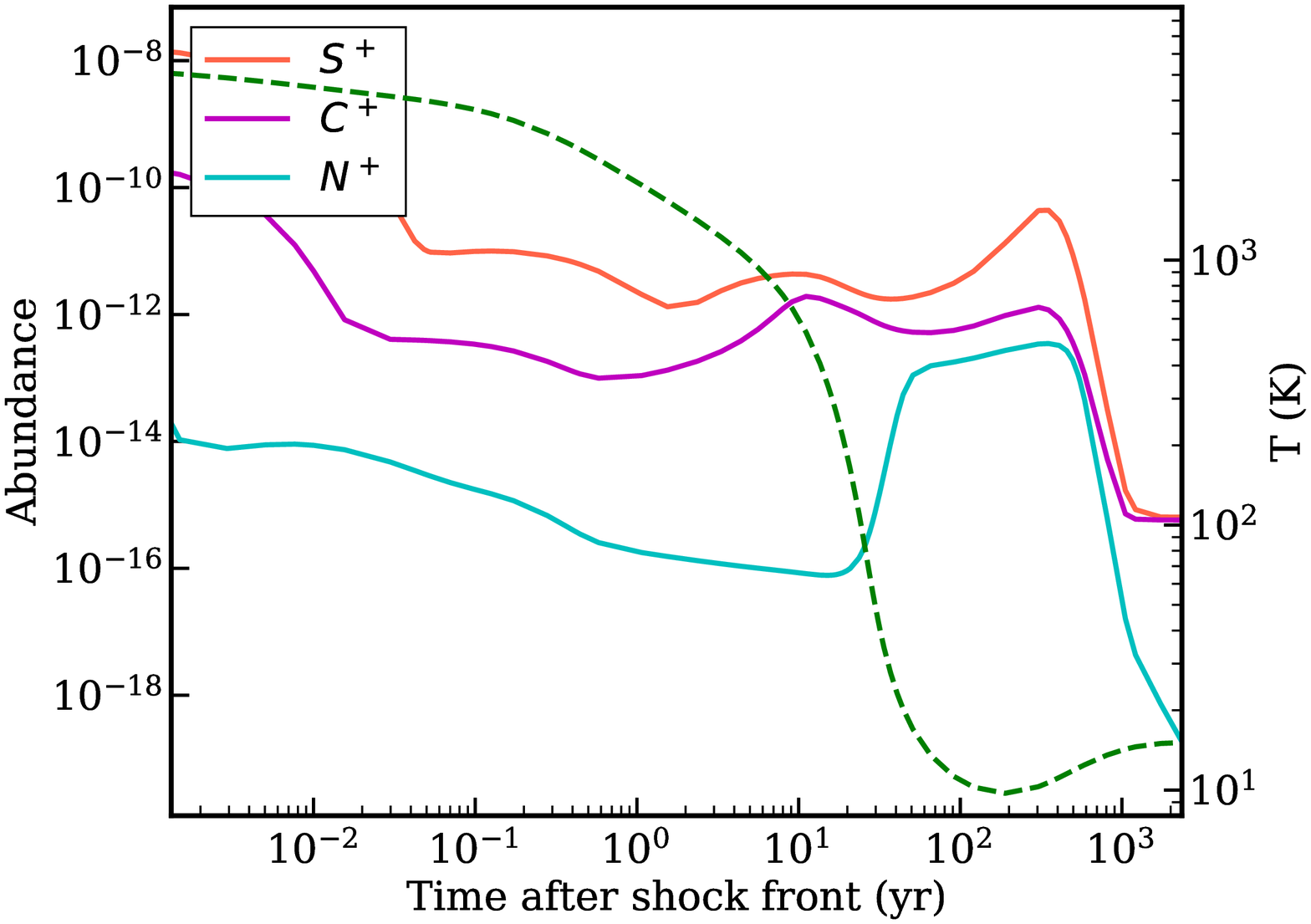}

     	\caption{ Left: Abundances of carbon-chain molecules from our single point gas-grain chemical model for cold/warm gas;  Right: Abundances of some elementary ions after the passage of a shock \citep{2012MNRAS.421.2786F,2015A&A...578A..63F}.
                  The dotted green line represents gas temperature. \label{fig_model}
     }
     \end{figure*}

\clearpage

\begin{table*}
\caption{Sources \label{table_source}}
\begin{tabular}{llllccccc}
\hline\hline
Name             & Ra(J2000) &  Dec(J2000) &  D(pc)    &    Notes        &   Reference      	 & Observation date\\
\hline
L1660              & 07:20:06.75& -24:02:20.9  &	1000	 & Molecular outflow     & \tablenotemark{a}& 2016.03.24	          	 \\
IRAS 20582+7724     & 20:57:10.6& +77:35:46	   &	200      & Molecular outflow, L1228 & \tablenotemark{b} &2016.01.25/26,2017,11,9 \\
L1221			    &	22:28:02.7&	+69:01:13	&	200	     & Molecular outflow  &	 \tablenotemark{c}	&	 2016.01.26,2017,11,9      \\
L1251A			     &	22:30:35.0&	+75:14:00	&	330	     & Molecular outflow  & \tablenotemark{d} & 2015.12.16, 2016.01.25 \\
Lup I-1          & 15:43:01.68& -34:09:08.9	   &	155	     & Molecular outflow,WCCC & \tablenotemark{e}  &	 2016.03.21    	\\
Lup I-2	           & 15:44:59.8 &-34:17:09	    &	155     & In Lup1 C6     & \tablenotemark{f}  &	2016.01.25	            	   \\
Lup I-3/4	      & 15:45:14.8	&-34:17:02.7&	155	  & In Lup1 C7         & \tablenotemark{f}	    &	2016.03.21       	      \\
Lup I-5	          & 15:45:03.80&-34:17:57.3	&	155	  & In Lup1 C6     & \tablenotemark{f}	    &	2016.03.21      	      \\
Lup I-6 	     & 15:42:52.4	&-34:07:54	&	155	 & Lupus-1A, in Lup1 C3    & \tablenotemark{g}	    &	 2016.01.25	            \\
Lup I-7/8/9        & 15:42:44.06&-34:08:30.4	&	155	   & In Lup1 C3      & \tablenotemark{f}     &	2016.03.24	                \\
Lup I-11         & 15:45:25.10&-34:24:01.8	&	155    & In Lup1 C8        & \tablenotemark{f}	    &	2016.03.24      \\	
\hline
\end{tabular}
\\
\tablenotetext{a}{\citet{2005ApJ...632..964L,1997A&A...324..263D,2004A&A...426..503W} and the references therein};
\tablenotetext{b}{\citet{2009AJ....137.3993D}};
\tablenotetext{c}{\citet{1991ApJ...377..510U}};
\tablenotetext{d}{\citet{2011ApJ...730L..18C}};
\tablenotetext{e}{\citet{2015A&A...584A..36G,2012MNRAS.419..238B, 2009ApJ...697..769S,2008A&A...480..785L}};
\tablenotetext{f}{\citet{2015A&A...584A..36G,2012MNRAS.419..238B,2008A&A...480..785L}};
\tablenotetext{g}{\citet{2015A&A...584A..36G,2012MNRAS.419..238B,2010ApJ...718L..49S,2008A&A...480..785L}}
\end{table*}

\begin{table*}
\caption{Observed line and telescope parameters \label{table_freq}}
\begin{tabular}{ccccccc}
\hline\hline
	Molecular &	Transition & freq.(MHz)  & $Log_{10}(A_{ij}\ s^{-1})$ &   $E_{up}(K)$ &  HPBWs (") \\
\hline
\hctn	&	J=2-1 F=1-1	&	18198.3745	    	&	-7.26550	&	 	 1.30990	 &	52	 \\
\nodata	&	J=2-1 F=3-2	&	18196.3104	    	&	-6.88533	&	 	 1.30995	 &	52	 \\
\nodata	&	J=2-1 F=2-1	&	18196.2169	    	&	-7.01030	&	 	 1.30980	 &	52	 \\
\nodata	&	J=2-1 F=1-0	&	18195.1364	    	&	-7.14068	&	 	 1.31003	 &	52	\\
\nodata	&	J=2-1 F=2-2	&	18194.9195	    	&	-7.48753	&		 1.30988	 &	52	 \\
\hcfn	&	J=6-5 F=7-6	&	15975.9831	    	&	-6.86356	&		 2.68359	 &	60	\\
\nodata	&	J=6-5 F=6-5	&	15975.9663	    	&	-6.87581	&		 2.68345	 &	60	 \\
\nodata	&	J=6-5 F=5-4	&	15975.9336	    	&	-6.87816	&		 2.68359	 &	60	\\
\hcsn	&	J=16-15	&	18047.9697	        	&	-6.11305	&		 7.36235	 &	53	 \\
\nodata	&	J=15-14	&	16919.9791	        	&	-6.19805	&		 6.49617	 &	56	 \\
\nodata	&	J=14-13	&	15791.9870	        	&	-6.28888	&		 5.68424	 &	60	 \\
\cts	&	J=3-2	&	17342.2564   	    	&	-6.44743	&		 1.66464	 &	55	 \\
\hline
\end{tabular}
\end{table*}

\begin{table*}
\centering
\caption{Observed Parameters \label{table_obs_par}}
\begin{tabular}{ccccccccccc}
\hline\hline
Lines	&	Transition	&	I20582	&	L1221	&	L1251A	 &	 LupusI-1	&	 LupusI-2	&	 LupusI-5	&	 LupusI-6	&	LupusI-7/8/9	&	 LupusI-11\\
\hline
&& \multicolumn{9}{c}{Part I \quad V$_\mathrm{lsr}$ (\kms)} \\
\hline
\hctn   &   J=2-1 F=1-1     &   -8.1(1) &   -4.53(4)  &   -3.92(3)  &    5.11(2)   &   4.88(3)   &   4.97(3)   &    5.13(2)   &   5.13(2)    &   4.38(2)     \\
&   J=2-1 F=3-2             &   -8.04(3)  &   -4.49(3)  &   -3.96(2)   &   5.10(2)    &  4.87(2)     &   4.95(2)    &   5.12(2)    &    5.14(2)   &   4.37(2)    \\
&   J=2-1 F=2-1             &   -8.06(4)  &   -4.51(4)  &   -3.95(2)   &   5.10(2)    &  4.86(2)     &   4.94(2)    &   5.12(2)    &    5.14(2)   &   4.37(2)    \\
&   J=2-1 F=1-0             &   -8.01(3)  &   -4.47(4)  &   -3.97(3)  &   5.09(2)    &  4.86(3)    &   4.94(3)   &   5.11(2)    &    5.14(2)   &   4.37(2)    \\
&   J=2-1 F=2-2             &   -8.12(6)  &   -4.46(5)  &   -3.96(4)  &   5.09(2)    &  4.84(3)    &   4.99(3)   &   5.14(2)    &    5.13(2)   &   4.37(2)    \\
\hcfn   &   J=6-5 F=7-6     &               &               &   -3.79(3)  &   5.08(2)    &    4.94(2)   &   5.09(2)    &   5.11(2)    &    5.14(2)   &   4.39(2)    \\
&   J=6-5 F=6-5             &   -8.00(4)  &   -4.59(5)  &               &   5.08(2)    &               &               &   5.11(2)    &   5.14(2)    &    4.39(2)   \\
&   J=6-5 F=5-4             &               &               &   -3.97(3)  &   5.09(4)    &   4.89(2)    &  4.96(2)     &   5.11(3)    &   5.14(2)    &    4.38(2)   \\
\hcsn   &   J=16-15         &   -7.95(7)  &               &   -4.00(3)  &   5.10(2)    &    4.87(3)  &   4.96(3)    &   5.11(2)    &   5.14(2)    &  4.39(2)    \\
&   J=15-14                 &               &   -4.49(9)  &   -3.96(4)  &   5.09(2)    &   4.83(3)   &   4.95(2)    &   5.10(3)    &   5.14(2)    &    4.39(2)   \\
&   J=14-13                 &   -7.73(8)  &               &   -4.09(6)  &   5.08(2)    &   4.82(3)   &   4.94(2)    &   5.12(2)    &   5.12(2)    &    4.37(2)   \\
\cts    &   J=3-2           &   -8.07(5)  &   -4.55(5)  &   -3.98(2)   &    5.12(3)  &   4.71(4)   &   4.94(3)   &   5.10(4)   &  5.14(3)    &   4.37(2)     \\
\hline
&&\multicolumn{9}{c}{Part II \quad $T_\mathrm{MB}$ (K)} \\
\hline
\hctn   &   J=2-1 F=1-1  &   0.05(2)  &  0.09(2)  &  0.37(8) &   1.5(2)   &   0.38(8)   &   0.7(2)   &   0.59(9) &  1.0(1)   &   0.88(9)    \\
&   J=2-1 F=3-2          &   0.28(2) &   0.30(2) &   1.57(8)&   5.1(2)   &   1.71(9)  &   2.9(2)  &   2.04(9)&    4.1(1) &   3.50(9)    \\
&   J=2-1 F=2-1          &   0.15(2) &   0.21(2) &   0.83(8) &   3.2(2)   &   1.08(9)  &   1.7(2)  &   1.39(9)&   2.6(1)  &  2.30(9)    \\
&   J=2-1 F=1-0          &   0.08(2)  &  0.13(2)  &  0.36(8) &   1.6(2)   &   0.48(9)   &   0.9(2)   &  0.82(9) &   1.3(1)  &    1.17(9)   \\
&   J=2-1 F=2-2          &   0.06(2)  &  0.08(2)  &  0.22(8) &   1.1(2)   &   0.39(9)   &   0.55(2)   &  0.63(9) &   1.2(1)  &    0.97(9)    \\
\hcfn   &   J=6-5 F=7-6  &           &           &   0.20(3) &   1.28(5)    &  0.53(4)   &   1.01(5)   &  0.62(5) &   0.93(5)    &   1.00(5)     \\
&   J=6-5 F=6-5          &   0.061(7)  &  0.057(8)  &           &   1.08(5)    &             &              &   0.47(5) &   0.79(5)    &  0.90(5) \\
&   J=6-5 F=5-4          &           &           &   0.14(3) &   0.92(5)     &   0.34(4)   &   0.61(6)    &   0.42(5) &   0.70(5)    &   0.81(5) \\
\hcsn   &   J=16-15      &   0.037(8)  &       &   0.08(4)  &   0.76(7)     &   0.28(5)   &   0.54(7)    &   0.40(5) &   0.55(5)    &   0.65(5)  \\
&   J=15-14              &       & 0.026(8)    &   0.10(4)  &   8.3(7)     &   0.26(5)   &    0.54(7)   &   0.30(5) &   0.53(5)    &   0.63(6) \\
&   J=14-13              &   0.030(8)  &       &   0.06(4)  &   0.75(7)     &   0.31(5)   &    0.65(7)   &   0.38(5) &   0.44(4)    &   0.72(6) \\
\cts    &   J=3-2        &   0.049(8)  &   0.036(8)  &   0.26(4) &   0.22(6)     &   0.11(4)   &   0.25(6)    &   0.10(5) &   0.13(4)    &   0.32(4)  \\
\hline
&&\multicolumn{9}{c}{Part III \quad FWHM (km s$^{-1}$)} \\
\hline
\hctn   &   J=2-1 F=1-1  &   1.0(2)   &   0.4(1) &   0.36(5) &       0.19(2) &   0.45(4)  &   0.33(4) &   0.22(3) &   0.18(3) &    0.22(3)\\
&   J=2-1 F=3-2          &   0.85(4)     &   0.74(4) &   0.34(2)  &   0.25(2) &  0.42(2)  &   0.42(2)  &   0.26(2)  &   0.21(2)  &   0.23(4)  \\
&   J=2-1 F=2-1          &   0.90(7)     &   0.80(6) &   0.36(3) &   0.25(2)  &   0.39(3)  &   0.41(3) &   0.24(3) &   0.20(2)  &   0.22(2)   \\
&   J=2-1 F=1-0          &   1.0(1)     &   0.54(8) &   0.35(3) &   0.23(3) &    0.44(4) &   0.41(3) &   0.20(3) &   0.18(3)  &    0.21(2)  \\
&   J=2-1 F=2-2          &   0.6(1)    &   0.5(1)&   0.40(5) &     0.24(3) &   0.40(4)  &   0.50(5) &   0.18(3) &   0.17(3) &   0.22(3)  \\
\hcfn   &   J=6-5 F=7-6  &               &           &           &   0.21(2)  &   0.67(3)  &    0.59(2)&   0.19(3)  &   0.18(2)  &   0.23(2)  \\
&   J=6-5 F=6-5          &   1.31(8)    &   0.8(1)&   0.56(3) &   0.27(3) &            &           &  0.19(3)  &   0.16(2)  &   0.18(2)  \\
&   J=6-5 F=5-4          &               &           &   0.39(4) &   0.23(2)  &   0.37(3)  &   0.39(3) &   0.18(2) &   0.16(2)  &   0.18(2)  \\
\hcsn   &   J=16-15      &   0.9(1)    &           &   0.35(7) &   0.29(2) &   0.43(4)  &    0.42(4)&   0.18(2) &   0.19(2)  &   0.25(2)  \\
&   J=15-14              &               &   0.64(3)&   0.31(5) &   0.26(2) &   0.46(4)  &   0.42(3) &    0.25(3)&   0.20(2) &   0.27(2) \\
&   J=14-13              &   0.6(2)    &           &   0.6(1)&   0.31(2) &   0.49(3)  &   0.41(3) &   0.22(3) &   0.25(2) &   0.26(2) \\
\cts    &   J=3-2        &   1.4(2)   &   0.8(1)&   0.32(3) &   0.24(3) &    0.54(2)  &  0.30(4)  &   0.3(1)&   0.35(5) &    0.24(3)\\
\hline
	&&\multicolumn{9}{c}{Part IV \quad Area (K km s$^{-1}$)} \\
\hline
\hctn	&	J=2-1 F=1-1	&	    0.041(8)	&	0.033(6)	&	0.14(2)	&	0.30(2)	&	 0.18(2)&	 0.25(2)	    &	0.14(1)	&	 0.19(1)	&	0.20(1)\\
     	&	J=2-1 F=3-2	&		0.26(1)	&	0.23(1)	&	0.57(2)	&	1.38(5)&	0.77(3)	&	 1.27(4)	&	0.55(2)	&	 0.90(3)	&	0.85(3)	 \\
    	&	J=2-1 F=2-1	&		0.13(1)	&	0.14(1)	&	0.32(1)	&	0.84(3)	&	0.45(2)	&	 0.78(3)	&	0.36(2)	&	 0.55(2)	&	0.54(2)	\\
     	&	J=2-1 F=1-0	&		0.066(7)	& 0.053(7)	&	0.13(1)	& 0.40(2)	&0.23(2)	&	0.41(3)	&	0.18(1)	&	 0.25(1)	&	0.27(1)	\\
     	&	J=2-1 F=2-2	&		0.031(6)	& 0.035(7)	&	0.10(1)	& 0.30(2)	& 0.17(1)	&	0.30(3)	&	0.12(1)	&	 0.21(1)	&	0.23(1)	\\
\hcfn	&	J=6-5 F=7-6	&			    &		    &	0.12(1)	&	0.28(1)	&	0.38(1)	&	 0.64(2)	&	0.12(1)	&	 0.18(1)	&	0.24(1)	\\
     	&	J=6-5 F=6-5	&		0.077(4)	&0.058(5)	&		    &	0.31(1)	&	     	&		        &	 0.096(6)	&	 0.013(6)	&	0.174(7)	\\
     	&	J=6-5 F=5-4	&	            &	    	&	0.041(4)	&	0.23(1)	&	0.13(1)	&	 0.25(1)	    &	 0.080(5)	&	 0.118(6)	&	0.158(6)	\\
\hcsn	&	J=16-15	    &		0.026(4)   &		    &	0.031(6)	&	0.24(1)	&	0.13(1)	&	 0.24(1)	&	0.077(6)	&	 0.109(6)	&	0.168(6)	\\
 	    &	J=15-14	    &		        &	0.013(3)   &	0.031(5)	&	0.23(1)	&	0.126(8)	&	 0.24(1)	    &	 0.078(7)	&	 0.110(6)	&	0.18(1)	\\
    	&	J=14-13	    &		0.014(3)	&		    &	0.038(7)	&	0.25(1)	&	0.16(1)	&	 0.28(1)   	&	 0.089(6)	&	 0.116(6)	&	0.199(9)	\\
\cts	&	J=3-2	    &		0.069(7)	&	0.024(3)	&	0.088(6)	&	0.054(7)	&	0.064(7)   &	 0.079(8)	    &	0.033(8)	&	 0.046(6)	&	0.083(6)	\\
\hline
\end{tabular}
\end{table*}

\begin{table*}
\centering
\caption{\hctn~HFS fitting \label{table_derived_hfs}}
\begin{tabular}{cccc}
\hline
Sources & $\tau$(\hctn) & Tex(\hctn) & T$_d$ \\
        &               &  K         & K\\
\hline
I20582&   $<$0.1   &   --  &    13.5(0.3)  \\
L1221&   0.28(0.19) &   14.02(5.11) &    15.1(0.2)  \\
L1251A &  $<$0.1  &  --  &    12.4(0.2)  \\
 LupusI-1 &   0.84(0.08) &   15.79(1.69) &    13.9(0.1)  \\
LupusI-2&   0.55(0.11)  &   9.52(1.32)  &    11.5(0.2)  \\
 LupusI-5&   0.72(0.13) &   11.93(1.71) &    11.2(0.1)  \\
 LupusI-6&   1.43(0.13) &  7.03(0.52) &    10.0(0.2)  \\
LupusI-7/8/9& 0.85(0.08)   & 14.50(1.33)   &    10.2(0.1)  \\
 LupusI-11&   1.07(0.08)&  11.39(0.77)&    11.9(0.8)  \\
\hline
\end{tabular}
\end{table*}

\begin{table*}
\centering
\caption{Column densities derived from each line component (in { units of} 10$^{12}$ cm$^{-2}$) \label{table_derived_par_detailed}}
\begin{tabular}{cccccccccccc}
\hline\hline
	Lines	&	Transition	&		I20582	&	L1221	&	L1251A	 &	 LupusI-1	 &	LupusI-2	&	 LupusI-5	&	 LupusI-6	&	LupusI-7/8/9	&	 LupusI-11\\
\hline
	&		&\multicolumn{9}{c}{Part I \quad Column density derived from each line component ($10^{12}$ cm$^{-2}$)} \\	
\hline
\hctn	&J=2-1 F=1-1	&21(4)	&20(3)	&69(6)	&157(8)	&83(6)	&112(11)	&57(4)	&80(4)	 &98(5)\\
	&J=2-1 F=3-2	&23.8(0.7)	&28(1)	&49.4(1.0)	&131(1)	&63(1)	&104(2)	&42.1(0.8)	 &69.7(0.8)	&72.1(0.9)\\
	&J=2-1 F=2-1	&22(1)	&30(1)	&52(1)	&148(3)	&69(2)	&118(3)	&50(1)	&79(1)	 &85(1)\\
	&J=2-1 F=1-0	&25(2)	&25(3)	&49(4)	&157(7)	&79(5)	&139(8)	&56(3)	&80(3)	 &95(3)\\
	&J=2-1 F=2-2	&16(3)	&22(4)	&46(5)	&156(10)	&78(6)	&135(12)	&52(3)	&89(4)	 &107(4)\\
\hcfn	&J=6-5 F=7-6	&	&	&7.8(0.6)	&35(1)	&24(1)	&40(1)	&12.7(0.5)	&19.1(0.5)	 &27.8(0.8)\\
	&J=6-5 F=6-5	&3.6(0.6)	&3.2(0.8)	&	&46(1)	&	&	&11.9(0.6)	&16.4(0.6)	 &23.6(0.8)\\
	&J=6-5 F=5-4	&	&	&6.7(0.7)	&40(1)	&20(1)	&39(1)	&11.8(0.7)	&17.5(0.7)	 &25.4(0.8)\\
\hcsn	&J=16-15	&1.1(0.2)	&	&1.3(0.2)	&10.0(0.3)	&5.2(0.3)	&9.8(0.4)	 &3.1(0.2)	&4.4(0.2)	 &6.9(0.2)\\
	&J=15-14	&	&0.7(0.2)	&1.4(0.2)	&10.2(0.4)	&5.4(0.3)	&10.2(0.4)	&3.3(0.3)	 &4.6(0.2)	&7.8(0.3)\\
	&J=14-13	&0.7(0.1)	&	&1.8(0.3)	&12.2(0.4)	&7.6(0.4)	&13.0(0.5)	&4.0(0.3)	 &5.2(0.2)	&9.3(0.3)\\
\cts	&J=3-2	&3.5(0.4)	&1.4(0.2)	&4.2(0.2)	&2.8(0.4)	&2.9(0.3)	&3.5(0.4)	 &1.4(0.3)	&2.0(0.3)	 &3.9(0.2)\\
\hline
	&		&\multicolumn{9}{c}{Part II \quad Column density for each specie ($10^{12}$ cm$^{-2}$)} \\	
\hline
\hctn	&	&21(2)	    &25(2)	&53(3)	&150(6)	&74(4)	&121(7)	&52(2)	&79(3)	&91(3)\\
\hcfn	&	&3.6(0.6)	&3.2(0.8)	&7.3(0.6)	&40(1)	&22(1)	&39(1)	&12.1(0.6)	 &17.7(0.6)	&25.6(0.8)\\
\hcsn	&	&0.9(0.2)	&0.7(0.1)	&1.5(0.3)	&10.8(0.4)	&6.1(0.3)	&11.0(0.4)	 &3.5(0.2)	&4.7(0.2)	 &8.0(0.3)\\
\cts	&	&3.5(0.4)	&1.4(0.2)	&4.2(0.2)	&2.8(0.4)	&2.9(0.3)	&3.5(0.4)	 &1.4(0.3)	&2.0(0.3)	 &3.9(0.2)\\
\hline
\end{tabular}
\end{table*}

\begin{table*}
\centering
\caption{Abundances (in unit $10^{-10}$) \label{table_derived_par}}
\begin{tabular}{ccccc}
\hline
Sources & \hctn & \hcfn & \hcsn & \cts \\
\hline
I20582&10(1)&1.7(0.3)&0.42(0.07)&1.7(0.2)\\
L1221&16(1)&2.2(0.4)&0.5(0.1)&1.01(0.08)\\
L1251A &26(1)&3.6(0.3)&0.7(0.1)&2.1(0.1)\\
 LupusI-1 &40(1)&11.0(0.3)&2.92(0.10)&0.76(0.10)\\
LupusI-2&25(1)&7.7(0.4)&2.1(0.1)&1.0(0.1)\\
 LupusI-5&43(2)&14.2(0.5) &3.9(0.2) &1.3(0.1) \\
 LupusI-6&13(1)&3.1(0.2)&0.89(0.06)&0.36(0.09)\\
LupusI-7/8/9&23(1) &5.0(0.2) &1.36(0.06) &0.56(0.07) \\
 LupusI-11&61(2)&17.1(0.5)&5.3(0.2)&2.6(0.2)\\
\hline
\end{tabular}
\end{table*}

 \begin{table*}
\centering
\caption{Abundance ratio \label{table_calculated_values}}
\begin{tabular}{ccccccc}
\hline
Sources & x(\hctn) & &x(\hcfn) && x(\cts) &\\
        & /x(\hcfn) &Average&   /x(\hcsn) &Average& /x(\hcsn)&Average\\
\hline
I20582&   5.8   & &         4.0  & &   5.0& \\
L1221&    7.3    & &        5.0  & &   2.0& \\
L1251A &  7.3   &  6.8 &       4.9  & 4.6 &  3.3 &3.4 \\
\hline
 LupusI-1 & 3,8  &  3.8 &      3.7  & 3.7  & 0.3  &0.3\\
\hline
LupusI-2&   3.3  &    &     3.6  & &   0.5 & \\
 LupusI-5&  3.1  &   &      3.5  &  &  0.3 & \\
 LupusI-6&   4.3 &    &     3.5  &  &  0.4 & \\
LupusI-7/8/9& 4.5   &  &    3.8  &   & 0.4 & \\
 LupusI-11&   3.6 &  3.8  &    3.2  & 3.5  &  0.5 &0.4 \\
\hline
\end{tabular}
\end{table*}

\clearpage

\appendix
\section{Dust Temperatures and Column Densities}   \label{sec_appendix}

		All the sources observed except L1660 have \textit{Herschel} data at 70,
		160, 250, 350, and 500 \um~in the
		Herschel Science Archive\footnotemark[3],
		where we extracted Level 2 or Level 2.5 images. These data,
		which are from the Gould Belt Survey and a few other Herschel
		key programs and open time programs, have resolutions of
		8.$\!\!\arcsec$4, 13.$\!\!\arcsec$5, 18.$\!\!\arcsec$1,
		24.$\!\!\arcsec$9 and 36.$\!\!\arcsec$4 at
		70, 160, 250, 350, and 500 \um~respectively\footnotemark[3]$^,$\footnotemark[4],
		and the corresponding pixel sizes are 3.$\!\!\arcsec$2,
		3.$\!\!\arcsec$2, 6.$\!\!\arcsec$0, 10.$\!\!\arcsec$0 and 14.$\!\!\arcsec$0.
		On the basis of these multi-bands far-IR data, we obtained
		dust temperatures and column densities.

		\footnotetext[3]{\url{http://www.cosmos.esa.int/web/herschel/science-archive}}
		\footnotetext[4]{\url{http://herschel.esac.esa.int/Docs/PACS/html/pacs_om.html}}
           \footnotetext[5]{\url{http://starlink.eao.hawaii.edu/starlink/WelcomePage}}
    \footnotetext[6]{\url{http://www.starlink.ac.uk/docs/sun255.htx/sun255ss4.html}}

    \subsection{Background Removal}

    Background/foreground emission was removed using the
    \textit{CUPID-findback} algorithm of the \textit{Starlink}
    suite\footnotemark[5] \citep{2014ASPC..485..391C}. The algorithm constructs the background
    iteratively from the original image. At first, a filtered form of the
    input data is produced by replacing every input pixel
    by the minimum of the input values within a rectangular box
    centered on the pixel. This filtered data is then filtered again,
    using a filter that replaces every pixel value by the maximum value
    in a box centered on the pixel. Then each pixel in this
    filtered data is replaced by the mean value in a box centered
    on the pixel. The same box size is used for the first three steps.
    The final background estimate is obtained via some corrections and
    iterations by comparisons with the initial input data. More details about
    the algorithm can be found on the online document of
    \textit{findback}\footnotemark[6]. As a key parameter, the box
    has been assigned to be the source size at 250 \micron~which was
    measured based on the area with emission higher than 50\% of the peak
    intensity for each target.

    \subsection{Flux Measurements}

\begin{figure*}
\centering
\includegraphics[width=110mm]{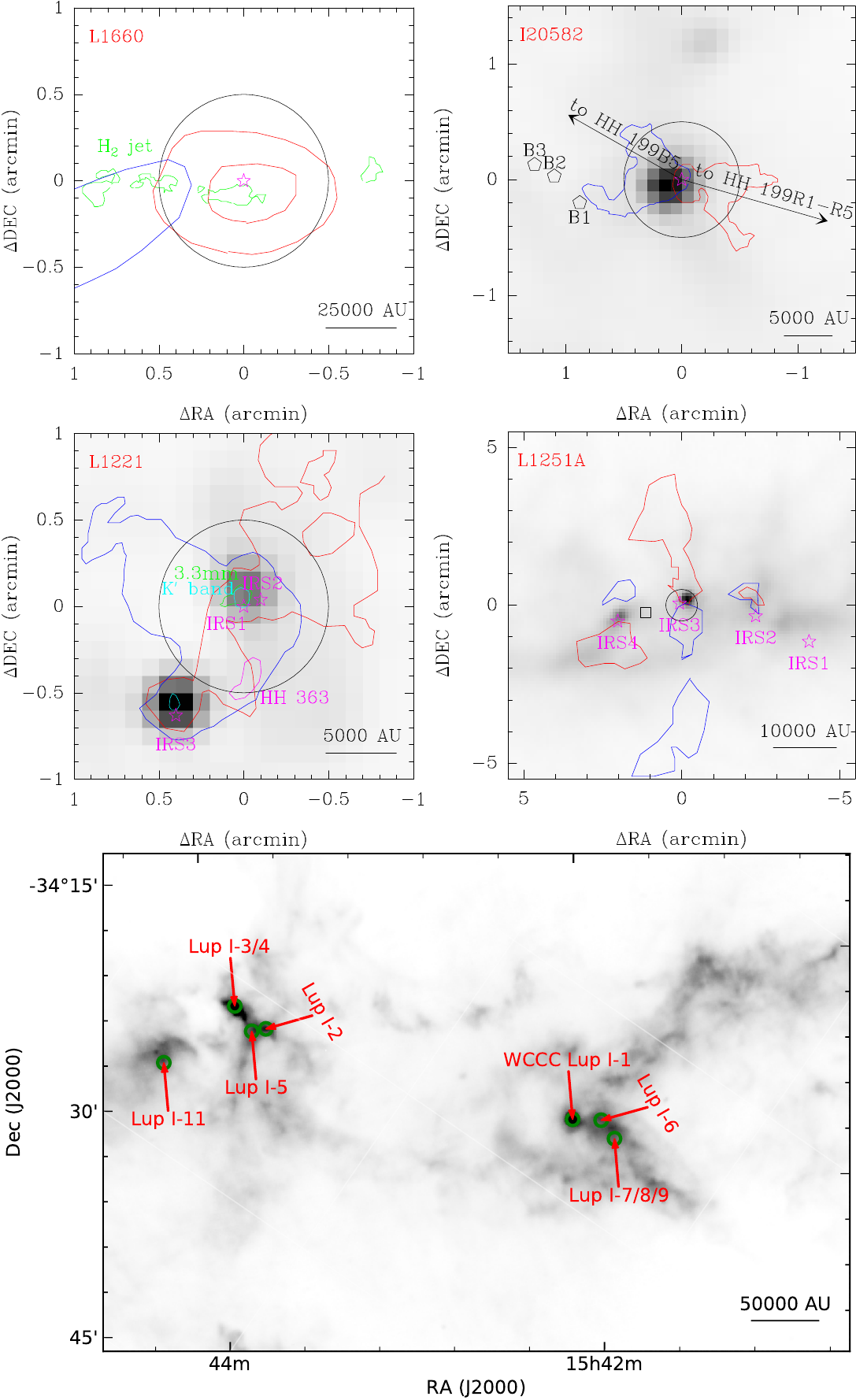}
\caption{Upper 4 panels show materials covered within a single beam (black circle) for outflow sources
L1660 \citep{1997A&A...324..263D,1988ApJ...327..350S}, I20582 \citep{2004ApJ...612..342A},
L1221 \citep{1997IAUS..182P..51A,2005ApJ...632..964L,2009ApJ...702..340Y} and
L1251A \citep{2010ApJ...709L..74L}.
Blue and red lobes of CO outflows are roughly sketched by blue and red solid lines.
The black square in the panel of L1251A denotes the point observed by \citet{2011ApJ...730L..18C}.
Bottom panel shows the locations of the observed Lupus I cores and the beams (green circles).  Background are Herschel 250 $\mu$m
maps \citep{2010A&A...518L.102A,2013A&A...549L...1R}.
\label{beams}}
\end{figure*}

    For each source, we firstly determined the source size based on the emission
    at 250 \micron~(\autoref{beams}) via fitting an ellipse to the
    region with emission higher than 50\% of the peak intensity around the
    target. Then fluxes at other bands were obtained by integrating
    the emission encompassed by the ellipse if the ellipse { was} larger than
    the beam. Otherwise fluxes referring to the beam centered at the source
    have been used.
    The source size at 250 \micron~and fluxes at Herschel bands are given in 2ed-7th
    column of \autoref{table_SED}.

    \subsection{Spectral Energy Distribution Fitting}

    The fluxes at \textit{Herschel} bands have been modeled using
    single temperature gray-bodies,
    \begin{equation}\label{eq-gb}
    S_\nu=B_\nu(T) (1-e^{-\tau_\nu})\Omega
    \end{equation}
    where the Planck function $B_\nu(T)$ is modified by { the} optical depth
    \begin{equation}
    \tau_\nu = \mu_\mathrm{H_2}m_\mathrm{H}\kappa_\nu N_\mathrm{H_2}/R_\mathrm{gd}.
    \end{equation}
    Here, $\mu_\mathrm{H_2}=2.8$
    is the mean molecular weight adopted from \citet{2008A&A...487..993K},
    $m_\mathrm{H}$ is the mass of a
    hydrogen atom, $N_\mathrm{H_2}$ is the column density, $R_\mathrm{gd}=100$ is
    the gas to dust ratio. The dust opacity $\kappa_\nu$
    can be expressed as a power law of frequency,
    \begin{equation}
    \kappa_\nu=5.9\left(\frac{\nu}{850~\mathrm{GHz}}\right)^\beta~\mathrm{cm^2g^{-1}}.
    \end{equation}
    with $\kappa_\nu(\mathrm{850~GHz})=5.9~\mathrm{cm^2g^{-1}}$ adopted
    from \citet{1994A&A...291..943O}.
    The $\Omega$ in \autoref{eq-gb} is
    the solid angle of the target. The
    free parameters are the dust temperature, dust emissivity index $\beta$
    and column density.

    The fluxes at 70 \um~were excluded from the SED fitting because emission at this
    wavelength may arise from a warmer dust component and a large fraction of the 70 \um~emission
    originates from very small grains, where the assumption of a single equilibrium
    temperature is not valid.

\begin{figure*}
\centering
\includegraphics[width=140mm]{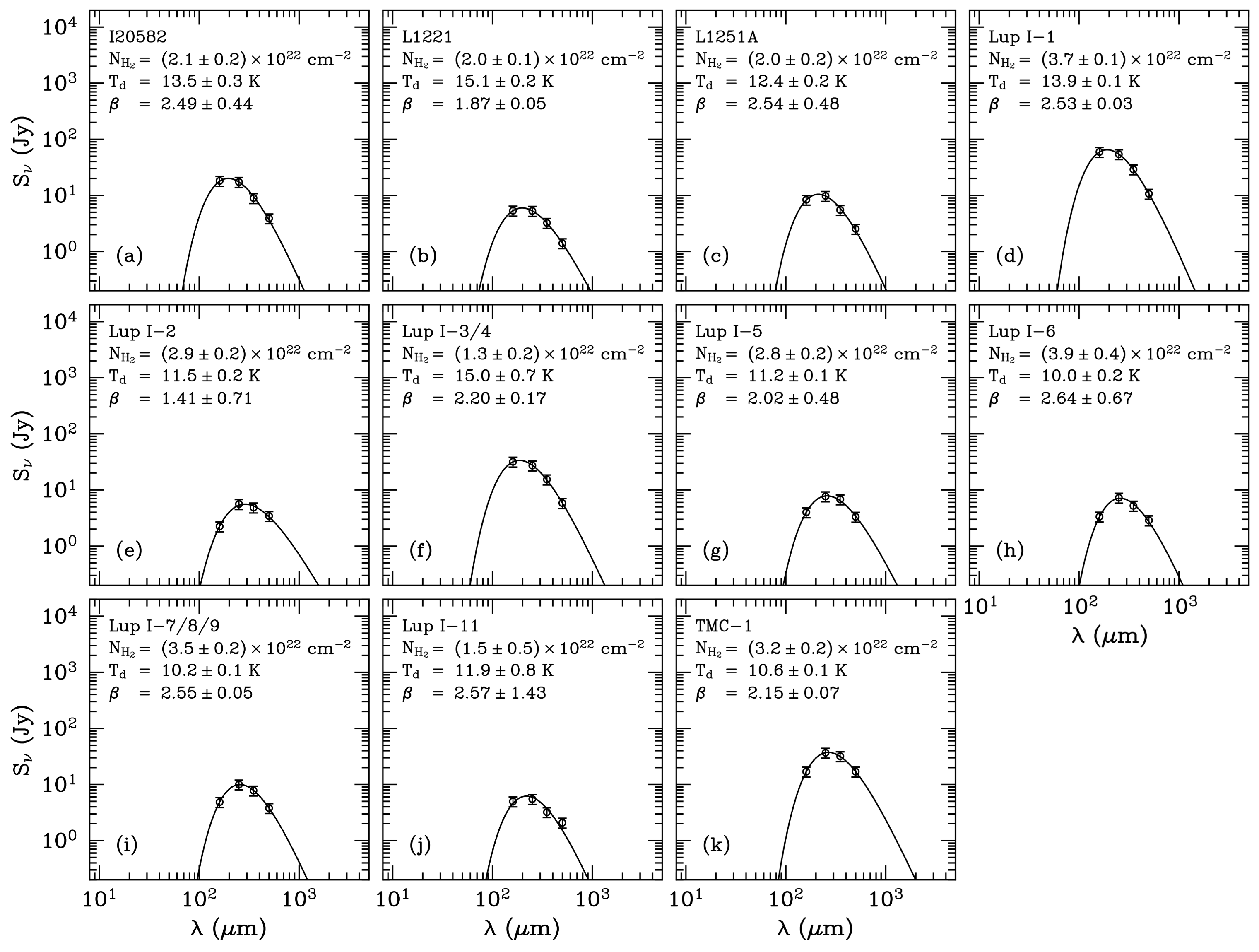}
\caption{SED of the observed sources. For comparisions the SED of TMC-1 is also presented. 
          The filled circles are the data points used for the SED fitting.
          The error bar of each point is plotted. \label{fig_SED}  }
\end{figure*}

    The fitting was performed with the Levenberg-Marquardt algorithm provided in
    the python package \textit{lmfit} \citep{2016ascl.soft06014N}.
    Fitted SED curves were shown in \autoref{fig_SED}.
    The resulting dust temperatures and column
    densities are listed in columns 8 and 9 of
    \autoref{table_SED}.

\begin{table*}
\caption{Dust Parameters \label{table_SED}}
\begin{tabular}{lrrrrrrrr}
\hline\hline
\colhead{Source} & \colhead{S$_{70}$}& \colhead{S$_{160}$}& \colhead{S$_{250}$}&
\colhead{S$_{350}$}& \colhead{S$_{500}$}& \colhead{Size$_{250}$} &\colhead{T$_\mathrm{dust}$} &\colhead{N$_\mathrm{H_2}$}\\
& \colhead{Jy} & \colhead{Jy} & \colhead{Jy} & \colhead{Jy} & \colhead{Jy} & \colhead{arcsec}
& \colhead{K} & \colhead{$10^{22}$ cm$^{-2}$} \\
\hline
 IRAS20582+7724    &    9.04      &    18.14   &      17.28   &    8.94      &  3.89   &    43.16 &    13.5(0.3)  & 2.1(0.2)  \\
          L1221    &     6.22     &    5.34    &     5.28     &    3.23      &   1.40  &    21.91 &    15.1(0.2)  & 2.0(0.1)  \\
         L1251A    &     1.75     &    8.35    &     9.77     &    5.50      &   2.53  &    40.67 &    12.4(0.2)  & 2.0(0.2)  \\
       Lupus1-1    &    21.49     &   59.44    &    54.10     &   29.07      &  10.67  &    55.00 &    13.9(0.1)  & 3.7(0.1)  \\
       Lupus1-2    &     0.19     &    2.24    &     5.61     &    4.86      &   3.44  &    36.00 &    11.5(0.2)  & 2.9(0.2)  \\
     Lupus1-3/4    &     8.86     &   31.73    &    27.13     &   15.37      &   5.83  &    57.96 &    15.0(0.7)  & 1.3(0.2)  \\
       Lupus1-5    &     0.25     &    4.00    &     7.67     &    6.79      &   3.32  &    42.69 &    11.2(0.1)  & 2.8(0.2)  \\
       Lupus1-6    &     0.04     &    3.34    &     7.26     &    5.22      &   2.87  &    43.00 &    10.0(0.2)  & 3.9(0.4)  \\
   Lupus1-7/8/9    &     0.00     &    4.86    &    10.00     &    7.79      &   3.80  &    50.42 &    10.2(0.1)  & 3.5(0.2)  \\
      Lupus1-11    &     0.32     &    3.31    &     5.50     &    3.21      &   2.08  &    39.20 &    11.9(0.8)  & 1.5(0.5)  \\
      TMC-1        &     0.14     &    16.95   &     36.78    &    31.92     &   16.96 &   100.00 &    10.6(0.1)  & 3.2(0.2)  \\
\hline
\end{tabular}\\
\end{table*}


\bsp
\label{lastpage}
\end{document}